\documentclass[aps,pre,twocolumn,superscriptaddress,showpacs,amsmath,amssymb]{revtex4-1}
\usepackage{graphics}

\begin{document}

\title{Recent advances and open challenges in percolation} 

\author{N. A. M.  Ara\'ujo}
\email{nmaraujo@fc.ul.pt}
\affiliation{Departamento de F\'{\i}sica, Faculdade de Ci\^{e}ncias,
Universidade de Lisboa, 1749-016 Lisboa, Portugal, and Centro de
F\'isica Te\'orica e Computacional, Universidade de Lisboa, 1749-016
Lisboa, Portugal}

\author{P. Grassberger}
\email{p.grassberger@fz-juelich.de}
\affiliation{JSC, FZ J\"ulich, D-52425 J\"ulich, Germany}

\author{B. Kahng}
\email{bkahng@snu.ac.kr}
\affiliation{Department of Physics and Astronomy, Seoul National University 151-747, Korea}

\author{K. J. Schrenk}
\email{kjs73@cam.ac.uk}
\affiliation{Computational Physics for Engineering Materials, IfB, ETH
Zurich, Wolfgang-Pauli-Strasse 27, CH-8093 Zurich, Switzerland and
Department of Chemistry, University of Cambridge, Lensfield Road,
Cambridge, CB2 1EW, United Kingdom}

\author{R. M. Ziff}
\email{rziff@umich.edu}
\affiliation{Center for the Study of Complex
Systems and Department of Chemical Engineering, University of Michigan,
Ann Arbor, Michigan 48109-2136 USA}

\begin{abstract}
Percolation is the paradigm for random connectivity and has been one of
the most applied statistical models. With simple geometrical rules a
transition is obtained which is related to magnetic models. This
transition is, in all dimensions, one of the most robust continuous
transitions known. We present a very brief overview of more than 60
years of work in this area and discuss several open questions for a
variety of models, including classical, explosive, invasion,
bootstrap, and correlated percolation.
\end{abstract}

\maketitle

\section{Introduction}\label{intro}
Percolation is a classic problem in statistical physics and, like a cat
with nine lives, never seems to die.  The study by its proper name began
with the work of engineer Simon Broadbent and mathematician John
Hammersley in the 1950's~\cite{Broadbent57}, inspired by the workings of
activated charcoal gas masks, but it was effectively already considered
by chemist Paul Flory in the early 1940's in his study of gelation in
polymers~\cite{Flory41a,Flory41b,Flory41c}.  The King's College London
group of Cyril Domb, which included Michael Fisher, John Essam, and M. F.
Skyes, did a great deal to popularize the percolation problem in the
physics community starting in the 1960's~\cite{Essam80}. In an early
paper, Fisher and Essam showed that the polymerization model of Flory
corresponds to percolation on the Bethe
lattice~\cite{Fisher61a,Fisher61}. There followed a great deal of work
by physicists, including Rushbrooke, Stanley, Coniglio, Halperin,
Herrmann, Stauffer, Aharony, Havlin, Duplantier,
etc~\cite{Stauffer79,Stauffer94,Sahimi94}. A major breakthrough was the
demonstration in 1969 by Fortuin and Kasteleyn \cite{Kasteleyn69}
that percolation is a limiting case of the general Potts model, which
includes the Ising model, and can be solved exactly. This paved the way
to many exact results in percolation, and allowed powerful
renormalization group ideas to be used~\cite{Cardy96}. Numerous variations of
percolation, such as invasion percolation, first-passage percolation,
directed percolation, and bootstrap percolation were introduced. Exact
``static" exponents in 2d were first proposed in the 1980's by den Nijs
\cite{denNijs79,Nienhuis80,Pearson80}, Pearson and others, but attempts to find exact
exponents for conductivity (the Alexander-Orbach
conjecture)~\cite{Alexander82}, the
backbone~\cite{Grassberger92,Grassberger99}, and the shortest path
(chemical distance)~\cite{Herrmann84} have not succeeded.

Finding percolation thresholds both exactly and by simulation has been
an enduring subject of research in this field (see \cite{Schrenk13} and
references therein), as well as the development of algorithms such as
those by Hoshen and Kopelman~\cite{Hoshen76}, by Leath~\cite{Leath76},
and by Newman and Ziff~\cite{Newman00}. Finding rigorous proofs of
exact thresholds and bounds has also been an enduring area of research
for mathematicians (Kesten~\cite{Kesten80}, Wierman~\cite{Wierman84},
Grimmett~\cite{Grimmett99}, Bollob\'as and
Riordan~\cite{Bollobas10}, etc.).  

The derivation of an exact formula for the crossing probability on a
rectangle by Cardy in 1992~\cite{Cardy92} (motivated by numerical work
of Langlands et al. \cite{Langlands92}) led to a great deal of  work on
universal crossing properties in two-dimensional system, such as those
by Pinson~\cite{Pinson94}, Watts~\cite{Watts96},
Simmons~\cite{Simmons07},  etc. Around 2000, Schramm developed the
Stochastic Loewner Evolution (SLE) theory, which was soon applied to
percolation hulls~\cite{Schramm00,Schramm01,Schramm09}. This caused once
again percolation to be an active area of research in mathematical
probability theory, and led to two Fields' medals, to Stanislav Smirnov
and Wendelin Werner~\cite{Smirnov01,Smirnov01b,Smirnov06}.  The results
include rigorous derivation of the static percolation exponents.
Another infusion of interest in percolation came from the surging field
of network theory, which goes back to the study of random and complete
graphs by Erd\H os and R\'enyi (1959), where the formation of a ``giant
component" is exactly analogous to percolation~\cite{Erdos60}, but was
revitalized by interest in small-world and scale-free networks.  In
2000, Newman and Moore found the critical point for a random graph in
the limit of large size, in which case the system is effectively a Bethe
lattice, and this result connects to the early work of Flory, Fisher and
Essam, but here with a general degree distribution~\cite{Moore00}. In
the field of random networks, the model of explosive percolation was
first introduced, by Achlioptas, D'Souza and
Spencer~\cite{Achlioptas09}, and this has been another problem that has
led to a wave of new interest in percolation. 

In this article, we present some perspective on open problems and
challenges in percolation. The field is so vast that we can only touch
on a subset of them, more aligned with our own interests.

\section{Classical uncorrelated
percolation}\label{classical.percolation}
The most basic question in percolation is the value of the threshold for
a given system.  Exact thresholds in two dimensions for the square,
triangular, honeycomb and related lattices (kagom\'e and (3-12$^2$) site
thresholds) were found by Sykes and Essam in 1963, using the
star-triangle transformation \cite{Sykes63}.  In 1984 Wierman
\cite{Wierman84} generalized this transformation to find the threshold
of the bow-tie lattice, and in 2006 Ziff and Scullard \cite{Ziff06}
showed that thresholds can be found for any lattice that can be
represented as a self-dual 3-hypergraph (that is, decomposed into
triangles that form a self-dual arrangement).  Just recently, Grimmett
and Manolescu \cite{Grimmett13} showed that thresholds can be found for
any lattice that can be represented geometrically as an isoradial graph,
yielding a broad new class of exact thresholds and providing a proof
\cite{Ziff12} of Wu's 1979 conjecture  \cite{Wu79} for the threshold of
the checkerboard lattice.  So the widely held belief that there are
just a handful of lattices where the threshold is known exactly is no
longer true.  However, many systems of long interest (such as site
percolation on the square and honeycomb lattices, and bond percolation
on the kagom\'e lattice) do not fit into the self-dual hypergraph or
isoradial forms and cannot presently be solved.  Is there another
approach that can give exact thresholds to these and other lattices?  Or
can a proof be made that thresholds for some lattices can never be found
in a closed form?  New methods to obtain very high precision estimates
of the threshold with unprecedented precision (over 12 digits for some
models) have been developed by Jacobsen and Scullard
\cite{Jacobsen13,Jacobsen14}, but none of these seem to suggest any
obvious closed-form values.  For higher dimensions, exact thresholds
would be very beneficial but seems unlikely.  In four dimensions, a
claim for an exact threshold of plaquettes has been made
\cite{Miyazima06}.

Likewise, there are no exactly known exponents for more dynamical
properties such as the shortest-path exponent $d_\mathrm{min}$, dynamic
exponent $z$, conductivity exponent $t$, spectral dimension, etc., even
in two dimensions (only the related exponent $g$ is known
\cite{Grassberger99b,Ziff99b}).  For directed percolation, which can be
interpreted as a dynamical form of percolation also related to the
contact process, no exact results are known for any dimension.  After
years of trying, it seems unlikely that these other exponents will ever
be found exactly, but perhaps a new breakthrough will be found.

Many questions relating to scaling functions and amplitude ratios were
looked at extensively in the 1980's, but several questions remain
unanswered, and with new computers and algorithms these problems are
worth revisiting~\cite{Ziff11c}. Amplitude ratios are useful for
confirming universality of different percolation models
\cite{Aharony97,deSouza11}.  Near the percolation threshold the size
distribution of percolation clusters satisfies the scaling form $n_s(p)
\sim A s^{-\tau} f(B(p - p_c)s^\sigma)$ where $A$ and $B$ are
system-dependent scale factors, while $\tau$ and $\sigma$ are universal
exponents and $f(z)$ is a universal function for a given dimensionality.
The universality of $f(z)$ implies the universality of amplitude ratios
(such as the ratio of the mean cluster size an $\epsilon$ above and
below $p_c$), which were studied extensively in the past
\cite{Privman91}.  With a precise determination of the scaling function,
many of these universal ratios should be able to be determined, but as
far as we know the scaling function has never been determined precisely
and used for this.  For the correlation-length $\xi$, the amplitude
ratio has been predicted to be exactly 2~\cite{Chayes89}, but
because of questions of how the correlation length should be measured,
it seems that this value has never been verified numerically.  In 2001,
Seaton \cite{Seaton01} predicted an exact amplitude ratio for
percolation, $R_\xi^+ =  [\alpha (2 - \alpha) (1 - \alpha) A]^{1/d}
\xi_0^+ =  [40/27 \sqrt 3]^{\frac 1 2} = 0.9248\ldots$ where $\alpha =
-2/3$,  $A^+$ is the coefficient of the singular part $|p - p_c|^{2 -
\alpha}$ of the free energy (number of clusters) and $\xi_0^+$ is the
coefficient to the divergence of the correlation length $\xi \sim
\xi_0(p_c - p)^\nu$ for $p < p_c$.  While this ratio agreed with
previous results by
Delfino and Cardy \cite{Delfino98}, questions about the correct
definition of the correlation function to use remains. 

A way to put the size distribution into a universal form is by
considering the enclosing area or volume; in that case the area of the
$n$-th ranked cluster satisfies Zipf's law $A_n\sim C/n$, with $C$ known
exactly in the case of two-dimensional enclosed area
\cite{Cardy03,Ziff99c}.  How this quantity behaves away from $p_c$ and
how that relates to regular scaling functions has not been detailed.
This same approach should apply to any critical system, including
directed percolation.

An early exact result in percolation is the density $n$ of clusters of
wet sites in bond percolation at the threshold, found by Temperley and
Lieb~\cite{Temperley71} for the square lattice and evaluated simply as
$(3 \sqrt{3}-5)/2$ \cite{Ziff97b}, and is also known for bond
percolation on the triangular lattice~\cite{Baxter78}.  Surprisingly,
while many things have been proven exactly for site percolation on the
triangular lattice, this quantity is known only numerically there: $n  =
0.0176255(5)$ \cite{Ziff97b}.  Can this quantity also be found exactly?

\section{Classical percolation with long-range correlated
disorder}\label{correlated}
Percolation theory and related models have been applied to study
transport and geometrical properties of disordered systems
\cite{Isichenko92,Sahimi93}. Frequently, such disorder is characterized
by a power-law long-ranged spatial correlation. This fact has motivated
studies of percolation models where the sites of the lattice are not
occupied independently, but instead with long-range spatial correlation,
in a process named \emph{correlated percolation}
\cite{Weinrib84,Prakash92,Isichenko92,Sahimi93,Schmittbuhl93,Sahimi94a,Du04,Makse95,Makse98,Sahimi98,Makse00,Araujo02,Araujo03,Sandler04,Schrenk13b}.
The qualitative picture that emerged is that, in the presence of
long-range correlations, percolation clusters become more compact and
their transport properties change accordingly. These findings have also
been confirmed by experimental studies of the transport properties of
clusters in correlated invasion percolation
\cite{Auradou99,Schmittbuhl00}.

The typical strategy to investigate correlated percolation is to work
with a landscape of random heights $h$, where $h(\mathbf{x})$ is the
height of the landscape at the lattice site at position $\mathbf{x}$
\cite{Weinrib84,Prakash92,Isichenko92,Schmittbuhl93,Kalda08,Kondev00}.
Recently, \emph{ranked surfaces} have been introduced, providing the
adequate framework to tackle this problem~\cite{Schrenk12,Schrenk13b}.
The ranked surface of a discrete landscape is constructed as follows:
One first ranks all sites in the landscapes according to their height,
from the smallest to the largest value. Then, a ranked surface is
constructed where each site has a number corresponding to its position
in the rank. The following percolation model can then be defined:
Initially, all sites of the ranked surface are unoccupied. The sites are
occupied one-by-one, following the ranking. At each step, the fraction
of occupied sites $p$ increases by the inverse of the total number of
sites in the surface. By this procedure, a configuration of occupied
sites is obtained, the properties of which depend on the landscape. For
example, if the heights are distributed uniformly at random, classical
percolation with fraction of occupied sites $p$ is obtained
\cite{Newman00,Newman01,Hu12}. To generate ranked surfaces with
power-law correlated disorder the Fourier filtering method is
used~\cite{Peitgen88,Prakash92,Lauritsen93,Makse96,Ballesteros99b,Malamud99,Oliveira11,Ahrens11,Morais11}.

The focus is usually on the dependence of percolation properties on how
the correlations decay with the spatial distance, typically
characterized by the Hurst exponent $H$. Schmittbuhl {\it et al.} have
shown that for $H>0$, there is no percolation
transition~\cite{Schmittbuhl93}. Instead, compact clusters grow and
merge in such a way that the size of the largest cluster grows linearly
with the fraction of occupied sites. By contrast, for $H<0$ the
percolation transition is critical and critical exponents can be found.
According to the extended Harris criterion, the critical exponents
should change with
$H$~\cite{Weinrib83,Weinrib84,Schmittbuhl93,Sandler04,Janke04b,Schrenk13b}.
It was found that the fractal dimension of the largest cluster, its
perimeter, as well as the dimension of its shortest path, backbone, and
red sites depend on
$H$~\cite{Prakash92,Schmittbuhl93,Kalda08,Mandre11,Schrenk13b}. Strong
dependence on $H$ is also found for the electrical conductivity exponent
of the largest cluster and the growth of bridge sites in the correlated
percolation model.  Schrenk {\it et al.} have proposed a functional
dependence of several exponents on the Hurst exponent
$H$~\cite{Schrenk13b}. While these relations were found as the simplest
rational expressions giving the considered critical exponents as
function of $H$, they do not have theoretical support. It is still an
open question if these expressions can be found by scaling arguments or
field theory.  Concerning the perimeter fractal dimension of the largest
cluster, it was found numerically that the duality relation of
Duplantier~\cite{Duplantier00} is fulfilled in the full range of Hurst
exponents~\cite{Schrenk13b}. A more theoretical argument supporting this
finding is also still missing.

\section{Variants of percolation}\label{variants.of.percolation}
``Classical'' (ordinary) percolation deals \cite{Stauffer94,Sahimi94}
with the situation where small bridges are either established or removed
randomly, irreversibly, and independently, until large scale
connectivity is either established or lost. Here, a ``bridge'' might be
a link or a node in a network, with regular lattices being very regular
class of special networks. Apart from this no other processes of
relevance go on, which means that each bridge is dealt with only once,
whence the randomness or the phenomenon can be either considered as {\it
frozen} (a view mostly taken in condensed matter) or 
as characterizing a stochastic contagion {\it process}, in
which case it is describing the spreading of some
epidemic~\cite{Grassberger83}, computer
malware, or public opinion. In the latter context, it is also called the
SIR (``susceptible-infected-removed'') epidemic model
\cite{Mollison77,Murray05,deSouza11}, to be contrasted to the SIS
(``susceptible-infected-susceptible'') model where bridges are removed
after a finite life time (of infectivity), and which corresponds to
directed percolation in space-time (the directness being that of the
time arrow). Both SIR and SIS (i.e., undirected and directed
percolation) show transitions between phases with and without long range
connectivity, which are continuous (``second order'') in the sense that
a suitable order parameter (which usually is the density of the largest
connected cluster) is zero at the transition point and increases
continuously (but of course with singular derivatives) as one enters the
supercritical (connected) phase.  Deviations from this classical model
can involve any of the basic ingredients:
\begin{enumerate}
\item ``Bridges'' (links, infections) themselves can already be {\bf
long ranged}. Models with power behaved long range interactions have
been studied in statistical physics since the late 1960's , when it was
realized that the 1-dimensional Ising model can have a non-trivial phase
transition if interactions are sufficiently long ranged, with a decay
$\sim 1/r^2$ being the turning point
\cite{Dyson69,Anderson70,Thouless69}. Very early, Thouless
\cite{Thouless69} argued that the transition in this model is of
Kosterlitz-Thouless \cite{Kosterlitz73} type, i.e. it is of infinite
order, the correlation length does not scale with a power of the
distance from the critical point, and there are generic power laws with
continuously varying exponents in the subcritical phase. The same should
be true for percolation. But his arguments were soon shown to be
incomplete \cite{Anderson70}. Indeed, when it was proven that in this
case the transition is first order (discontinuous) \cite{Aizenman86},
the predictions of Kosterlitz seemed to be obsolete. But recent
simulations \cite{Grassberger13} show that there are indeed continuously
varying critical exponents (in the supercritical phase), and that the
correlation length diverges exponentially.  Related to this is a model
by Boettcher {\it et al.} \cite{Boettcher11}, where long range links are
not established in a completely random fashion. Rather, in the Boettcher
{\it et al.} model all long range bonds must have length $2^k,
k=1,2,3,\ldots$, bonds with length $2^k$ connect only to sites with
coordinates $i\equiv 0 \;\; {\rm mod }\;\; 2^k$, and they are all
established with the same probability $p$. This leads to the same
average density of long range links as in the random model with
$P(r)\sim 1/r^2$, and both models show very similar behavior. How similar
they are in details remains to be seen. 
\item In two dimensions, there is no
similarity with the Kosterlitz-Thouless transition, but there is a
long-standing controversy. When increasing the power in the contact
probability $\sim 1/r^{\sigma}$, there comes a point where the critical
behavior changes from a regime with $\sigma$-dependent critical
exponents to ordinary (short-range) percolation. The controversy deals
with the question when this happens. Different field theoretical
arguments suggest that it either happens when also the scaling of the
supercritical process changes, or already for smaller $\sigma$.
Extensive simulations (both for percolation and for the Ising model,
see~\cite{Linder08,Luijten02}) seemed to have settled this in favor of
the latter option, but very recent simulations suggest that neither is
right and things might be more
complicated~\cite{Grassberger13b,Picco12,Blanchard12}.  
\item Bridges can be established in a {\bf not entirely random} way. In
particular, when establishing a new bridge, one may take two random (and
not yet established) ``candidates,'' and establish that candidate that
leads to less long range connectivity (i.e., to a smaller cluster). This
was proposed by Achlioptas {\it et al.} \cite{Achlioptas09}, who termed
it ``explosive percolation" and claimed it to have a discontinuous
transition, where the transition is delayed so long that it finally
takes place in a sudden, ``explosive'' way. This paper has led to
several similar models, some of which actually do have first-order
transitions (see e.g.  \cite{Araujo10,Cho13}) --- although the original
Achlioptas model does not \cite{daCosta10,Riordan11,Grassberger11}.
Notice that the choice between the two candidates introduces a severe
non-locality which goes far beyond anything that can be described by a
local field theory. It is presumably for this reason that the Achlioptas
model seems to show (there is not yet any analytic proof) finite size
scaling that is qualitatively different from that expected from the
renormalization group \cite{Grassberger11}. This family of models is
discussed in detail in Sec.~\ref{explosive}.
\item Bridges can be established in a {\bf partly synchronized} way. In
ordinary bond percolation, links in a graph are randomly selected one
after the other. In ``agglomerative percolation''
\cite{Bizhani11,Son11,Christensen11} this is replaced by the rule that a
cluster is selected randomly, and then {\it all} links connected to it
are established immediately. On non-bipartite lattices in two and three
dimensions this is in the universality class of ordinary percolation,
but not in one-dimensional lattices \cite{Son11}, on a class of trees
\cite{Bizhani11}, and on Erd\H{o}s-R\'enyi graphs \cite{Bizhani11b}.
Notice that this model also introduces some non-locality (albeit much
weaker than in the Achlioptas model), if very large clusters are chosen
during late stages of the process. A special feature happens on
bipartite lattices like the square and simple cubic lattice
\cite{Christensen11,Lau12}. There, at the critical point also a $Z_2$
symmetry is spontaneously broken, which leads at least in two dimensions
to a new universality class.
\item In the standard scenario, bridges are newly established in a
pre-existing ``virtual'' network e.g., in bond percolation all nodes and
all possible links are already defined, while in site percolation the
new sites are chosen from a pre-existing set and all links are already
established. In contrast to this, Callaway {\it et al.}
\cite{Callaway01} considered the case where links are established in a
{\bf growing network}.  More precisely, each time a node is added to the
existing network, also a link is added with probability $\delta < 1$. In
contrast to the Barab\'asi-Albert model \cite{Barabasi99}, where the new
links are added so that nodes with higher degree get attached
preferentially, new links are added completely at random. Again this
leads to a Kosterlitz-Thouless type transition when $\delta$ passes
through a critical value, but now the transition is (as the original KT
transition) of infinite order instead of being discontinuous.
\item The establishment of bridges can be a {\bf cooperative} effect.
This seems to be the most interesting case.  Models of this type have
been studied as {\it bootstrap} percolation (or as the closely related
{\it k-core percolation}) since the 1970's \cite{Chalupa79,Adler91},
while co-epidemics (see below) are a major health hazard.
\begin{itemize} 
\item In {\bf bootstrap percolation} with index $m$ one considers a
connected cluster on some graph, and removes recursively all nodes which
have less than $m$ neighbors in the cluster. For $m=1$ nothing is done,
as a connected graph has always at least one neighbor. For $m=2$ all
leaves are removed in a first step. This might turn some nodes, which
originally had $\geq 2$ neighbors, into leaves. They are removed in a
second step. If this again turns nodes into leaves, they are removed
again, etc. In this way one ends up either with an empty cluster, or
with a cluster where all nodes have $\geq 2$ neighbors. In $k-$core
percolation one starts with the cluster which contains all nodes and
keeps at the end only the largest connected component, i.e. one ends up
finally with the largest connected subgraph in which all nodes have
degree at least $k$.  Bootstrap percolation can show either continuous
or discontinuous transitions, depending on the type of networks (lattice
or random), and on the dimensionality when the network is a regular
lattice. For details see \cite{Adler91}.
\item Instead of deleting ``poor'' nodes, let us now consider the
opposite process of {\bf complex contagion} \cite{Dodds04,Dodds05}. Here
nodes are attached to a growing cluster and nodes with more neighbors
in the cluster are more likely to be attached. A possible application is
political opinion spreading.  We assume that no person is so eloquent
that (s)he alone is able to convince his or her neighbor. But if there
is already a small group of {\it early adapters}, then the combined
argumentation of several of them is able to convince others with
probability $q$. When people get convinced if and only if they have
$\geq m$ convinced neighbors, this leads exactly to bootstrap
percolation. More interesting, however, is the case where the chance
$q_m$ to be finally convinced (or infected, or attached to the cluster,
depending on the application) depends non-trivially on $m$, the number
of neighbors in the attacking cluster.  Take e.g. site percolation on a
square lattice, where any site is incorporated with probability $p$ into
a growing cluster, if it has at least one infected neighbor. In this
case $q_m = p$ for all $m\ge 1$. In bond percolation, on the other hand,
each site has the same chance $p$ to convince (or infect) any not yet
convinced (or healthy) neighbor. A site with $m$ infected neighbors
succumbs then with probability $q_m =p+(1-p)p+ \ldots (1-p)^{m-1}p =
1-(1-p)^m$.
In general we expect $q_m$ (the probability that $m$ neighbors together
will lead to infection) to grow with $m$.  If this growth is moderate
(as e.g. in bond percolation) we expect the transition to be continuous
and in the universality class of ordinary percolation. This is no longer
true if $q_m$ grows too fast. In that case one finds first-order
transitions, where the epidemic (political opinion, fashion, computer
virus, ...) either does not take off at all or takes off explosively. In
between the first and second order regimes there is a tricritical point
whose properties on finite-dimensional lattices where studied by Jansen
{\it et al.} by field theoretic methods \cite{Janssen04}.  
\item For random
sparse (e.g. Erd\H{o}s-R\'enyi) networks the tricritical point is
particularly easy to find \cite{Dodds04,Bizhani12}: If $q_1>0$, the
transition from second to first order happens exactly when $q_2=2q_1$,
independent of its degree distribution.
\item One important physical application of this model is to critically pinned
driven interfaces in isotropic random media \cite{Barabasi95}. It is
well known that such interfaces are often self-affine (i.e., show
nontrivial scaling), while the bulk behind them is not fractal (as it
would be for critical percolation). Thus in this problem the phase
transition is actually {\it hybrid}, i.e. it combines a jump in the
order parameter (the density of the infected phase) with non-trivial
scaling related to the interface. The fact that any first order
transition in $k-$core percolation is actually hybrid was stressed in
particular in \cite{Goltsev06}. 
\item Notice that interfaces in the present model have in general overhangs
(as in real isotropic media, but in contrast to the standard model
\cite{Barabasi95,Doussal02} used to describe them). This has important
consequences.  First of all, such interfaces in 1+1 dimension are always
in the ordinary percolation universality class \cite{Bizhani12}, as
conjectured already in \cite{Drossel98} (see, however,
\cite{Zhou10,Qin12}).  This is related to the proof of \cite{Aizenman89}
that there cannot be first-order transitions in 2-dimensional random
media. Secondly, it means that there will always be some holes left in a
growing cluster, and these holes in general will show non-trivial
scaling at the pinning point (as required by its hybrid nature).
Finally, it might imply that the point where the zero temperature random field
Ising model changes from being hysteresis-free to having hysteresis is
not, as claimed in~\cite{Sethna06}, a critical point with upper critical
dimension $d_c=6$, but a tricritical point with $d_c=5$.  The field
theoretical predictions of \cite{Janssen04} for this tricritical point
were not well confirmed by the simulations of \cite{Bizhani12} for
$d=3$, but they were very well confirmed for $d=4$, where they are
supposed to be much more precise.
\item Instead of cooperativity between attacking neighbors, we can also
have cooperativity between different {\bf modes of attack}. This refers
in particular to {\bf cooperative co-epidemics} or {\bf syndemics}
\cite{Singer09}. Take e.g. the examples of the 1918 Spanish flu and TB
\cite{Brundage08,Oei12}, or of HIV and a host of other diseases like
diabetes, hepatitis, TB and others \cite{Sulkowski08,Sharma05}. In these
cases one disease weakens the victim so much that the victim falls prey to
the other one. A simple SIR type model for this with permanently
increased susceptibility for the other disease was developed in
\cite{Chen13c}. Simulations of this model on random graphs and on
lattices \cite{Cai14} show a host of different behaviors, depending on
the topology of the graph but also on the detailed of the implementation
(and, in one case, even on the initial conditions). But whenever a
first-order transition is found, it is in general hybrid.
\end{itemize}
\item The last model we shall discuss here are {\bf interdependent
networks}. Consider e.g.  a country like Italy with towns connected by a
network of electric power lines and another network of computer
connections. The latter are needed for controlling the power stations,
while the former are needed to provide power to the computers. If some
node fails, then this can lead to a huge cascade of failures as happened
indeed in Italy on September 28, 2003. As shown by Buldyrev {\it et al.}
in a seminal paper \cite{Buldyrev10}, this failure would have been a
first-order transition, if it had happened on a locally loopless network
like e.g. a sparse Erd\H{o}s-R\'enyi network. But, as shown in
\cite{Son11}, it still would be a continuous transition on a
2-dimensional regular lattice. Since Italy is 2-dimensional but not
quite regular, it is unclear what should apply in this case. But in any
case, in a series of papers (see e.g. \cite{Parshani10,Gao11,Son12})
several other types of interdependencies were studied. In all cases with
first-order transitions, these are indeed again hybrid.
\end{enumerate}
Finally, let us point out a possible connection between interdependent
networks and cooperative co-epidemics. People have livers and lungs, and
both can become sick. Someone with a liver disease will be more likely
to get also a lung infection and vice versa. Let us now assume that
liver and lung infections are both lethal. In this extreme case we are
dealing exactly with two interdependent networks as described in
\cite{Buldyrev10}. Details of this correspondence have not yet been
worked out.

\section{Explosive percolation}\label{explosive}
Conventionally percolation transitions are known to be continuous.
Recently, however, interest in discontinuous percolation transitions
(DPT) has been triggered and boosted by i) the explosive percolation
model~\cite{Achlioptas09} and ii) the cascading failure model in
interdependent networks~\cite{Buldyrev10}.  Actually the subject of DPT
was initiated by the $k$-core percolation
model~\cite{Chalupa79,Adler91,Bollobas84,Dorogovtsev06} a long
time ago. However, the mechanism of the DPT arising in the $k$-core
model was unusual, so that further studies of the DPT had not proceeded
very much. The two models i) and ii) were designed
for DPTs; however, the governing processes to the PT and their
mechanisms are different: cluster coagulation and fragmentation
processes for the model i) and ii), respectively. Moreover, the order
parameters are also different. Thus, we need to discuss the two problems
separately, and find common features in a unified framework. 

We first discuss the DPT driven by cluster merging processes.  The
explosive percolation (EP) model was introduced motivated by a
mathematical invention, the so-called Achlioptas process: at each step,
we take the one (among multiple options) which is the optimal one for the
formation of a given target pattern. To generate a DPT in random graph,
at each step, two pairs of nodes that are not connected yet are selected
at random, and one pair of nodes among them is taken and connected by a
new link, which is the one that produces a smaller size of connected
cluster than the size of the cluster produced by the other option. Then,
the growth of large clusters is suppressed and clusters with medium
sizes are abundant. As the number of added links is increased, such
medium-sized clusters merge and create a macroscopic-sized cluster,
which emerges drastically at a delayed percolation threshold. 

Following this pioneering paper~\cite{Achlioptas09}, many related models
have been
introduced~\cite{Ziff09,Cho09,Friedman09,DSouza10,Araujo10,Nagler11,Ziff13,Cho14}.
One of noticeable issues addressed in following works was that the EP
transition in random graph is not indeed discontinuous but continuous in
the thermodynamic limit. Actually, due to extremely slow convergence to
asymptotic behavior as the system size is increased, whether the EP is
indeed continuous or discontinuous was hard to be determined numerically
and became a controversial issue. This claim was addressed firstly based
on numerical results for a specific model~\cite{daCosta10}; however,
since the claim was not firmly supported by analytic solution, the claim had
not been firmly accepted first time. Later, based on the numerically
observed fact that the average size of medium-sized cluster is not
sufficiently large~\cite{Friedman09}, a mathematical
proof~\cite{Riordan11} was presented, which is the following: the number
of clusters participated to generate a macroscopic-sized cluster is not
sub-extensive to system size in order of magnitude, and thus, it cannot
bring out a DPT, but a continuous PT.  

The EP transition in Euclidean space has been also extensively
studied~\cite{Ziff09,Ziff10}. However, whether the percolation
transition is discontinuous or continuous had not been clarified,
either. In spite of such extensive studies, understanding of the EP
transition in Euclidean space and on random graph in a unified scheme
had been absent. For the continuous PT case, the Erd\H{o}s and R\'enyi
model on random graph was regarded as the mean-field model of the
ordinary percolation model. To resolve this goal for the DPT case, a
model called the spanning cluster-avoiding (SCA) model was
introduced~\cite{Cho13}.  In this model, the target pattern in the
Achlioptas process is taken as a spanning cluster, which is actually
standard in Euclidean space. This model used previous results, the
scaling behavior of the number of bridge bonds above a percolation
threshold. Here the bridge bonds are those that would form a spanning
cluster if occupied~\cite{Schrenk12}.  Using the scaling formula for the
bridge bonds, the percolation thresholds of the SCA model were
calculated analytically for various numbers of potential links in the
Achlioptas process. This analytic result leads to the following
conclusion: the EP transition can be either continuous or discontinuous,
depending on the number of multiple options, if the spatial dimension is
less than the upper critical dimension, and is always continuous
otherwise. Subsequently, it was concluded that the transition of the
ordinary EP model is continuous as a mean-field solution of the SCA
model. 

Further intriguing and fundamental questions remain open. For example,
the DPT occurring in the SCA model is rather trivial in the sense that
the order parameter increases drastically all the way to unity at a
trivial percolation threshold, which is equal to the system size in the
thermodynamic limit. This behavior is similar to the one of the DPT
governed by the Smoluchowski coagulation equation with the reaction
kernel $k_ik_j\sim(ij)^\omega$ with $0\leq\omega<0.5$~\cite{Cho10}.
Therefore, it would be intriguing to study a non-trivial DPT model in
which the order parameter is increased all the way to a finite value
less than unity at a finite threshold, and gradually beyond that
point~\cite{Bohman04,Chen11,Chen11b,Schrenk11b,Panagiotou11,Zhang12,Chen13,Boettcher11}.
Investigation of the origin of such a non-trivial DPT would be also
interesting. Moreover, understanding of other discontinuous phase
transitions in non-equilibrium systems such as synchronization
transitions~\cite{Gomez-Gardenes11} and jamming
transitions~\cite{Echenique05} in the perspective of the DPT under
cluster merging processes would be interesting.

Next we discuss the DPT occurring in inter-dependent networks. In this
DPT, the order parameter is taken as the fraction of nodes in the
mutually connected largest component, which is not the same as the
standard ones. The model associated with this DPT was introduced to
study cascading-failure dynamics in the inter-dependent systems. Even
though the original model was understood analytically, a simpler
analytic solution was derived from the viewpoint of epidemic
spreading~\cite{Son12}. In this perspective, the DPT can be understood
as the emergence of a mutually connected giant cluster in cluster
merging processes. Thus, it would be interesting to compare the
mechanism of such a DPT with that of the EP model. This investigation
would provide a clue to enhance the robustness of inter-dependent
networks. 

Percolation transitions have served as a platform for understanding
phase transitions in non-equilibrium systems. Likewise, the theoretical
framework developed to understand the origin of the DPTs occurring in
the models i) and ii) is anticipated to serve as a basis for further
research on drastic phase transitions in non-equilibrium complex
systems. 

\section{Retention capacity and watersheds of landscapes}\label{related}
Percolation has also been applied to study the properties of real and
artificial landscapes. A landscape is typically represented as a digital
elevation map (DEM), which consists of a two-dimensional array of
regular cells (sites) to which average heights can be associated. By
mapping the DEM to a ranked surface, it is possible to identify the
sequence of flooded sites when water is dripped over the landscape,
filling it from the valleys to the mountains, and letting the water flow
out through the open boundaries~\cite{Schrenk12}. In this spirit, one
can ask what the maximum volume of water is that can be retained by the
landscape. This problem can indeed be mapped to standard
percolation~\cite{Knecht11,Baek11}. For example, consider the simplest
case of a two-level random landscape of $L\times L$ sites, with a fraction
$p$ sites of unity level and $1-p$ of level zero. The water retention of
such a landscape is $R_2^{(L)}=L^2(p-P_\infty)$, where $P_\infty(p)$ is
the fraction of sites belonging to the percolation
cluster~\cite{Knecht11}. Since this cluster touches the borders, the water
that falls on its sites flows out of the landscape. In the general case
of a landscape with equal number of $n$ levels, it was
argued~\cite{Knecht11} that its retention $R_n^{(L)}$ can be expressed
as,
\begin{equation}
R_n^{(L)}=\sum_{i=1}^{n-1}R_2^{(L)}(i/n) \ \ .
\end{equation}
This expression was also shown to hold for correlated
landscapes~\cite{Schrenk14}. Some exact expressions have been found for
the retention capacity of finite (and very small) lattice
sizes~\cite{Knecht11,Schrenk14} but, a general analytic expression is
still missing. 

In hydrology it is also important to identify the watershed lines dividing
the landscape into different drainage basins. When the landscape is
flooded from the valleys lakes are formed. As the level of water rises,
those lakes start to merge and form even larger lakes. If one merges the
lakes under the constraint that no lake percolates, i.e., no lake
connects two predefined opposite boundaries of the landscape, one ends
up with only two lakes separated by the main watershed
line~\cite{Fehr09}. For random uncorrelated landscapes this line is a
fractal of dimension $1.2168\pm0.0005$~\cite{Fehr11c} and its
statistical properties are consistent with SLE$_\kappa$, with
$\kappa=1.734\pm0.005$~\cite{Daryaei12}. For correlated landscapes, the
fractal dimension of the watershed line decreases with the Hurst
exponent~\cite{Fehr11c,Fehr11}. The determination of the watershed line
can be mapped to a percolation problem where sites are sequentially
occupied (according to their rank in the ranked surface) under the
constraint that percolation is suppressed. Such description reveals a
tricritical point at a critical fraction of occupied (flooded) sites and
unveils how several seemingly unrelated physical models (e.g., optimal
path, optimal path cracks, and polymers in strongly disordered media)
fall into the same universality class~\cite{Schrenk12}. The
generalization of ranked percolation in three dimensions provides the
framework to determine the effective shares when different companies or
nations extract either oil, gas, or water, from the same porous
formation~\cite{Schrenk12b}.

\section{Directed percolation}
Let us add finally two comments on models related to directed
percolation.

One deals with the (in-)famous ``pair contact process with diffusion''
(PCPD) in one dimension of space. For a long time this was believed to
have continuously varying critical exponents, and yet two recent
simulations, Ref.~\cite{Schram12} and Ref.~\cite{Park12}, still come to
opposite conclusions.  While the former claim that it is in the directed
percolation universality class, the second rules this out.

The second deals with the ``parity conserving branching-annihilating
random walk'' (pcBARW)~\cite{Marro98}, which mainly differs from
directed percolation and the contact process by preserving the
``parity'' $P(t) = (-1)^{N(t)}$, where $N(t)$ is the number of particles
at time $t$.  When starting with two initial particles (i.\ e. in the
``even" sector), simulations (the most recent and careful ones being
in~\cite{Park13}) suggest that the exponent $\eta$ defined by $N(t) \sim
t^\eta$ is exactly equal to zero. This cries out for a theoretical
explanation, but so far none seems in sight.

\section{Final remarks}
Percolation is a vast subject with more than 80 thousand papers
published over the last 60 years according to the ISI Web of Knowledge,
and about one paper posted per day on the arXiv related to percolation.
Thus, instead of a comprehensive review (which would be an epic journey
and impossible in the space available), we decided to rather give a
flavor of this fascinating and active field and offer a glimpse at the
extensive list of still open questions, following our own interests. We
foresee many more years of interesting findings, constantly raising new
questions.

\begin{acknowledgments}
We acknowledge financial support from: (NA) the Portuguese Foundation
for Science and Technology (FCT), Contracts No. IF/00255/2013 and No.
EXCL/FIS-NAN/0083/2012; (BK) the Korean NRF, Grant No. 2010-0015066; and
(KJS) the Swiss National Science Foundation, Grant No. P2EZP2-152188. PG
thanks Golnoosh Bizhani for interesting discussion and collaboration on
cooperative percolation.
\end{acknowledgments}

\bibliography{percolation}

\begin{thebibliography}{181}%
\makeatletter
\providecommand \@ifxundefined [1]{%
 \@ifx{#1\undefined}
}%
\providecommand \@ifnum [1]{%
 \ifnum #1\expandafter \@firstoftwo
 \else \expandafter \@secondoftwo
 \fi
}%
\providecommand \@ifx [1]{%
 \ifx #1\expandafter \@firstoftwo
 \else \expandafter \@secondoftwo
 \fi
}%
\providecommand \natexlab [1]{#1}%
\providecommand \enquote  [1]{``#1''}%
\providecommand \bibnamefont  [1]{#1}%
\providecommand \bibfnamefont [1]{#1}%
\providecommand \citenamefont [1]{#1}%
\providecommand \href@noop [0]{\@secondoftwo}%
\providecommand \href [0]{\begingroup \@sanitize@url \@href}%
\providecommand \@href[1]{\@@startlink{#1}\@@href}%
\providecommand \@@href[1]{\endgroup#1\@@endlink}%
\providecommand \@sanitize@url [0]{\catcode `\\12\catcode `\$12\catcode
  `\&12\catcode `\#12\catcode `\^12\catcode `\_12\catcode `\%12\relax}%
\providecommand \@@startlink[1]{}%
\providecommand \@@endlink[0]{}%
\providecommand \url  [0]{\begingroup\@sanitize@url \@url }%
\providecommand \@url [1]{\endgroup\@href {#1}{\urlprefix }}%
\providecommand \urlprefix  [0]{URL }%
\providecommand \Eprint [0]{\href }%
\providecommand \doibase [0]{http://dx.doi.org/}%
\providecommand \selectlanguage [0]{\@gobble}%
\providecommand \bibinfo  [0]{\@secondoftwo}%
\providecommand \bibfield  [0]{\@secondoftwo}%
\providecommand \translation [1]{[#1]}%
\providecommand \BibitemOpen [0]{}%
\providecommand \bibitemStop [0]{}%
\providecommand \bibitemNoStop [0]{.\EOS\space}%
\providecommand \EOS [0]{\spacefactor3000\relax}%
\providecommand \BibitemShut  [1]{\csname bibitem#1\endcsname}%
\let\auto@bib@innerbib\@empty
\bibitem [{\citenamefont {Broadbent}\ and\ \citenamefont
  {Hammersley}(1957)}]{Broadbent57}%
  \BibitemOpen
  \bibfield  {author} {\bibinfo {author} {\bibfnamefont {S.~R.}\ \bibnamefont
  {Broadbent}}\ and\ \bibinfo {author} {\bibfnamefont {J.~M.}\ \bibnamefont
  {Hammersley}},\ }\href {\doibase 10.1017/S0305004100032680} {\bibfield
  {journal} {\bibinfo  {journal} {Proc. Cambridge Philos. Soc.}\ }\textbf
  {\bibinfo {volume} {53}},\ \bibinfo {pages} {629} (\bibinfo {year}
  {1957})}\BibitemShut {NoStop}%
\bibitem [{\citenamefont {Flory}(1941{\natexlab{a}})}]{Flory41a}%
  \BibitemOpen
  \bibfield  {author} {\bibinfo {author} {\bibfnamefont {P.~J.}\ \bibnamefont
  {Flory}},\ }\href@noop {} {\bibfield  {journal} {\bibinfo  {journal} {J. Am.
  Chem. Soc.}\ }\textbf {\bibinfo {volume} {63}},\ \bibinfo {pages} {3083}
  (\bibinfo {year} {1941}{\natexlab{a}})}\BibitemShut {NoStop}%
\bibitem [{\citenamefont {Flory}(1941{\natexlab{b}})}]{Flory41b}%
  \BibitemOpen
  \bibfield  {author} {\bibinfo {author} {\bibfnamefont {P.~J.}\ \bibnamefont
  {Flory}},\ }\href@noop {} {\bibfield  {journal} {\bibinfo  {journal} {J. Am.
  Chem. Soc.}\ }\textbf {\bibinfo {volume} {63}},\ \bibinfo {pages} {3091}
  (\bibinfo {year} {1941}{\natexlab{b}})}\BibitemShut {NoStop}%
\bibitem [{\citenamefont {Flory}(1941{\natexlab{c}})}]{Flory41c}%
  \BibitemOpen
  \bibfield  {author} {\bibinfo {author} {\bibfnamefont {P.~J.}\ \bibnamefont
  {Flory}},\ }\href@noop {} {\bibfield  {journal} {\bibinfo  {journal} {J. Am.
  Chem. Soc.}\ }\textbf {\bibinfo {volume} {63}},\ \bibinfo {pages} {3096}
  (\bibinfo {year} {1941}{\natexlab{c}})}\BibitemShut {NoStop}%
\bibitem [{\citenamefont {Essam}(1980)}]{Essam80}%
  \BibitemOpen
  \bibfield  {author} {\bibinfo {author} {\bibfnamefont {J.~W.}\ \bibnamefont
  {Essam}},\ }\href {\doibase 10.1088/0034-4885/43/7/001} {\bibfield  {journal}
  {\bibinfo  {journal} {Rep. Prog. Phys.}\ }\textbf {\bibinfo {volume} {43}},\
  \bibinfo {pages} {833} (\bibinfo {year} {1980})}\BibitemShut {NoStop}%
\bibitem [{\citenamefont {Fisher}\ and\ \citenamefont
  {Essam}(1961)}]{Fisher61a}%
  \BibitemOpen
  \bibfield  {author} {\bibinfo {author} {\bibfnamefont {M.~E.}\ \bibnamefont
  {Fisher}}\ and\ \bibinfo {author} {\bibfnamefont {J.~W.}\ \bibnamefont
  {Essam}},\ }\href@noop {} {\bibfield  {journal} {\bibinfo  {journal} {J.
  Math. Phys.}\ }\textbf {\bibinfo {volume} {2}},\ \bibinfo {pages} {609}
  (\bibinfo {year} {1961})}\BibitemShut {NoStop}%
\bibitem [{\citenamefont {Fisher}(1961)}]{Fisher61}%
  \BibitemOpen
  \bibfield  {author} {\bibinfo {author} {\bibfnamefont {M.~E.}\ \bibnamefont
  {Fisher}},\ }\href {\doibase 10.1063/1.1703746} {\bibfield  {journal}
  {\bibinfo  {journal} {J. Math. Phys.}\ }\textbf {\bibinfo {volume} {2}},\
  \bibinfo {pages} {620} (\bibinfo {year} {1961})}\BibitemShut {NoStop}%
\bibitem [{\citenamefont {Stauffer}(1979)}]{Stauffer79}%
  \BibitemOpen
  \bibfield  {author} {\bibinfo {author} {\bibfnamefont {D.}~\bibnamefont
  {Stauffer}},\ }\href@noop {} {\bibfield  {journal} {\bibinfo  {journal}
  {Phys. Rep.}\ }\textbf {\bibinfo {volume} {54}},\ \bibinfo {pages} {1}
  (\bibinfo {year} {1979})}\BibitemShut {NoStop}%
\bibitem [{\citenamefont {Stauffer}\ and\ \citenamefont
  {Aharony}(1994)}]{Stauffer94}%
  \BibitemOpen
  \bibfield  {author} {\bibinfo {author} {\bibfnamefont {D.}~\bibnamefont
  {Stauffer}}\ and\ \bibinfo {author} {\bibfnamefont {A.}~\bibnamefont
  {Aharony}},\ }\href@noop {} {\emph {\bibinfo {title} {Introduction to
  Percolation Theory}}},\ \bibinfo {edition} {2nd}\ ed.\ (\bibinfo  {publisher}
  {Taylor and Francis},\ \bibinfo {address} {London},\ \bibinfo {year}
  {1994})\BibitemShut {NoStop}%
\bibitem [{\citenamefont {Sahimi}(1994{\natexlab{a}})}]{Sahimi94}%
  \BibitemOpen
  \bibfield  {author} {\bibinfo {author} {\bibfnamefont {M.}~\bibnamefont
  {Sahimi}},\ }\href@noop {} {\emph {\bibinfo {title} {Applications of
  \mbox{Percolation} \mbox{Theory}}}}\ (\bibinfo  {publisher} {Taylor and
  Francis},\ \bibinfo {address} {London},\ \bibinfo {year} {1994})\BibitemShut
  {NoStop}%
\bibitem [{\citenamefont {Kasteleyn}\ and\ \citenamefont
  {Fortuin}(1969)}]{Kasteleyn69}%
  \BibitemOpen
  \bibfield  {author} {\bibinfo {author} {\bibfnamefont {P.~W.}\ \bibnamefont
  {Kasteleyn}}\ and\ \bibinfo {author} {\bibfnamefont {C.~M.}\ \bibnamefont
  {Fortuin}},\ }\href@noop {} {\bibfield  {journal} {\bibinfo  {journal} {J.
  Phys. Soc. Japan (Suppl.)}\ }\textbf {\bibinfo {volume} {26}},\ \bibinfo
  {pages} {11} (\bibinfo {year} {1969})}\BibitemShut {NoStop}%
\bibitem [{\citenamefont {Cardy}(1996)}]{Cardy96}%
  \BibitemOpen
  \bibfield  {author} {\bibinfo {author} {\bibfnamefont {J.}~\bibnamefont
  {Cardy}},\ }\href@noop {} {\emph {\bibinfo {title} {Scaling and
  Renormalization in Statistical Physics}}}\ (\bibinfo  {publisher} {Cambridge
  University Press},\ \bibinfo {address} {Cambridge},\ \bibinfo {year}
  {1996})\BibitemShut {NoStop}%
\bibitem [{\citenamefont {{den Nijs}}(1979)}]{denNijs79}%
  \BibitemOpen
  \bibfield  {author} {\bibinfo {author} {\bibfnamefont {M.~P.~M.}\
  \bibnamefont {{den Nijs}}},\ }\href {\doibase 10.1088/0305-4470/12/10/030}
  {\bibfield  {journal} {\bibinfo  {journal} {J. Phys. A}\ }\textbf {\bibinfo
  {volume} {12}},\ \bibinfo {pages} {1857} (\bibinfo {year}
  {1979})}\BibitemShut {NoStop}%
\bibitem [{\citenamefont {Nienhuis}\ \emph {et~al.}(1980)\citenamefont
  {Nienhuis}, \citenamefont {Riedel},\ and\ \citenamefont
  {Schick}}]{Nienhuis80}%
  \BibitemOpen
  \bibfield  {author} {\bibinfo {author} {\bibfnamefont {B.}~\bibnamefont
  {Nienhuis}}, \bibinfo {author} {\bibfnamefont {E.~K.}\ \bibnamefont
  {Riedel}}, \ and\ \bibinfo {author} {\bibfnamefont {M.}~\bibnamefont
  {Schick}},\ }\href {\doibase 10.1088/0305-4470/13/6/005} {\bibfield
  {journal} {\bibinfo  {journal} {J. Phys. A}\ }\textbf {\bibinfo {volume}
  {13}},\ \bibinfo {pages} {L189} (\bibinfo {year} {1980})}\BibitemShut
  {NoStop}%
\bibitem [{\citenamefont {Pearson}(1980)}]{Pearson80}%
  \BibitemOpen
  \bibfield  {author} {\bibinfo {author} {\bibfnamefont {R.~B.}\ \bibnamefont
  {Pearson}},\ }\href {\doibase 10.1103/PhysRevB.22.2579} {\bibfield  {journal}
  {\bibinfo  {journal} {Phys. Rev. B}\ }\textbf {\bibinfo {volume} {22}},\
  \bibinfo {pages} {2579} (\bibinfo {year} {1980})}\BibitemShut {NoStop}%
\bibitem [{\citenamefont {Alexander}\ and\ \citenamefont
  {Orbach}(1982)}]{Alexander82}%
  \BibitemOpen
  \bibfield  {author} {\bibinfo {author} {\bibfnamefont {S.}~\bibnamefont
  {Alexander}}\ and\ \bibinfo {author} {\bibfnamefont {R.}~\bibnamefont
  {Orbach}},\ }\href@noop {} {\bibfield  {journal} {\bibinfo  {journal} {J.
  Physique (Paris) Lett.}\ }\textbf {\bibinfo {volume} {43}},\ \bibinfo {pages}
  {625} (\bibinfo {year} {1982})}\BibitemShut {NoStop}%
\bibitem [{\citenamefont {Grassberger}(1992)}]{Grassberger92}%
  \BibitemOpen
  \bibfield  {author} {\bibinfo {author} {\bibfnamefont {P.}~\bibnamefont
  {Grassberger}},\ }\href {\doibase 10.1088/0305-4470/25/22/015} {\bibfield
  {journal} {\bibinfo  {journal} {J. Phys. A}\ }\textbf {\bibinfo {volume}
  {25}},\ \bibinfo {pages} {5867} (\bibinfo {year} {1992})}\BibitemShut
  {NoStop}%
\bibitem [{\citenamefont {Grassberger}(1999{\natexlab{a}})}]{Grassberger99}%
  \BibitemOpen
  \bibfield  {author} {\bibinfo {author} {\bibfnamefont {P.}~\bibnamefont
  {Grassberger}},\ }\href {\doibase 10.1016/S0378-4371(98)00435-X} {\bibfield
  {journal} {\bibinfo  {journal} {Physica A}\ }\textbf {\bibinfo {volume}
  {262}},\ \bibinfo {pages} {251} (\bibinfo {year}
  {1999}{\natexlab{a}})}\BibitemShut {NoStop}%
\bibitem [{\citenamefont {Herrmann}\ \emph {et~al.}(1984)\citenamefont
  {Herrmann}, \citenamefont {Hong},\ and\ \citenamefont
  {Stanley}}]{Herrmann84}%
  \BibitemOpen
  \bibfield  {author} {\bibinfo {author} {\bibfnamefont {H.~J.}\ \bibnamefont
  {Herrmann}}, \bibinfo {author} {\bibfnamefont {D.~C.}\ \bibnamefont {Hong}},
  \ and\ \bibinfo {author} {\bibfnamefont {H.~E.}\ \bibnamefont {Stanley}},\
  }\href {\doibase 10.1088/0305-4470/17/5/008} {\bibfield  {journal} {\bibinfo
  {journal} {J. Phys. A}\ }\textbf {\bibinfo {volume} {17}},\ \bibinfo {pages}
  {L261} (\bibinfo {year} {1984})}\BibitemShut {NoStop}%
\bibitem [{\citenamefont {Schrenk}\ \emph
  {et~al.}(2013{\natexlab{a}})\citenamefont {Schrenk}, \citenamefont
  {Ara\'ujo},\ and\ \citenamefont {Herrmann}}]{Schrenk13}%
  \BibitemOpen
  \bibfield  {author} {\bibinfo {author} {\bibfnamefont {K.~J.}\ \bibnamefont
  {Schrenk}}, \bibinfo {author} {\bibfnamefont {N.~A.~M.}\ \bibnamefont
  {Ara\'ujo}}, \ and\ \bibinfo {author} {\bibfnamefont {H.~J.}\ \bibnamefont
  {Herrmann}},\ }\href {\doibase 10.1103/PhysRevE.87.032123} {\bibfield
  {journal} {\bibinfo  {journal} {Phys. Rev. E}\ }\textbf {\bibinfo {volume}
  {87}},\ \bibinfo {pages} {032123} (\bibinfo {year}
  {2013}{\natexlab{a}})}\BibitemShut {NoStop}%
\bibitem [{\citenamefont {Hoshen}\ and\ \citenamefont
  {Kopelman}(1976)}]{Hoshen76}%
  \BibitemOpen
  \bibfield  {author} {\bibinfo {author} {\bibfnamefont {J.}~\bibnamefont
  {Hoshen}}\ and\ \bibinfo {author} {\bibfnamefont {R.}~\bibnamefont
  {Kopelman}},\ }\href {\doibase 10.1103/PhysRevB.14.3438} {\bibfield
  {journal} {\bibinfo  {journal} {Phys. Rev. B}\ }\textbf {\bibinfo {volume}
  {14}},\ \bibinfo {pages} {3438} (\bibinfo {year} {1976})}\BibitemShut
  {NoStop}%
\bibitem [{\citenamefont {Leath}(1976)}]{Leath76}%
  \BibitemOpen
  \bibfield  {author} {\bibinfo {author} {\bibfnamefont {P.~L.}\ \bibnamefont
  {Leath}},\ }\href {\doibase 10.1103/PhysRevB.14.5046} {\bibfield  {journal}
  {\bibinfo  {journal} {Phys. Rev. B}\ }\textbf {\bibinfo {volume} {14}},\
  \bibinfo {pages} {5046} (\bibinfo {year} {1976})}\BibitemShut {NoStop}%
\bibitem [{\citenamefont {Newman}\ and\ \citenamefont {Ziff}(2000)}]{Newman00}%
  \BibitemOpen
  \bibfield  {author} {\bibinfo {author} {\bibfnamefont {M.~E.~J.}\
  \bibnamefont {Newman}}\ and\ \bibinfo {author} {\bibfnamefont {R.~M.}\
  \bibnamefont {Ziff}},\ }\href {\doibase 10.1103/PhysRevLett.85.4104}
  {\bibfield  {journal} {\bibinfo  {journal} {Phys. Rev. Lett.}\ }\textbf
  {\bibinfo {volume} {85}},\ \bibinfo {pages} {4104} (\bibinfo {year}
  {2000})}\BibitemShut {NoStop}%
\bibitem [{\citenamefont {Kesten}(1980)}]{Kesten80}%
  \BibitemOpen
  \bibfield  {author} {\bibinfo {author} {\bibfnamefont {H.}~\bibnamefont
  {Kesten}},\ }\href@noop {} {\bibfield  {journal} {\bibinfo  {journal}
  {Commun. Math. Phys.}\ }\textbf {\bibinfo {volume} {74}},\ \bibinfo {pages}
  {41} (\bibinfo {year} {1980})}\BibitemShut {NoStop}%
\bibitem [{\citenamefont {Wierman}(1984)}]{Wierman84}%
  \BibitemOpen
  \bibfield  {author} {\bibinfo {author} {\bibfnamefont {J.~C.}\ \bibnamefont
  {Wierman}},\ }\href {\doibase 10.1088/0305-4470/17/7/020} {\bibfield
  {journal} {\bibinfo  {journal} {J. Phys. A}\ }\textbf {\bibinfo {volume}
  {17}},\ \bibinfo {pages} {1525} (\bibinfo {year} {1984})}\BibitemShut
  {NoStop}%
\bibitem [{\citenamefont {Grimmett}(1999)}]{Grimmett99}%
  \BibitemOpen
  \bibfield  {author} {\bibinfo {author} {\bibfnamefont {G.~R.}\ \bibnamefont
  {Grimmett}},\ }\href@noop {} {\emph {\bibinfo {title} {Percolation}}},\
  \bibinfo {series} {Grundlehren der mathematischen Wissenschaften}, Vol.\
  \bibinfo {volume} {321}\ (\bibinfo  {publisher} {Springer},\ \bibinfo {year}
  {1999})\BibitemShut {NoStop}%
\bibitem [{\citenamefont {Bollob\'as}\ and\ \citenamefont
  {Riordan}(2010)}]{Bollobas10}%
  \BibitemOpen
  \bibfield  {author} {\bibinfo {author} {\bibfnamefont {B.}~\bibnamefont
  {Bollob\'as}}\ and\ \bibinfo {author} {\bibfnamefont {O.}~\bibnamefont
  {Riordan}},\ }in\ \href@noop {} {\emph {\bibinfo {booktitle} {An Irregular
  Mind}}},\ \bibinfo {series} {Bolyai Society Mathematical Studies},
  Vol.~\bibinfo {volume} {21}\ (\bibinfo  {publisher} {Springer},\ \bibinfo
  {address} {Berlin},\ \bibinfo {year} {2010})\BibitemShut {NoStop}%
\bibitem [{\citenamefont {Cardy}(1992)}]{Cardy92}%
  \BibitemOpen
  \bibfield  {author} {\bibinfo {author} {\bibfnamefont {J.~L.}\ \bibnamefont
  {Cardy}},\ }\href {\doibase 10.1088/0305-4470/25/4/009} {\bibfield  {journal}
  {\bibinfo  {journal} {J. Phys. A}\ }\textbf {\bibinfo {volume} {25}},\
  \bibinfo {pages} {L201} (\bibinfo {year} {1992})}\BibitemShut {NoStop}%
\bibitem [{\citenamefont {Langlands}\ \emph {et~al.}(1992)\citenamefont
  {Langlands}, \citenamefont {Pichet}, \citenamefont {{Ph. Pouliot}},\ and\
  \citenamefont {{Saint-Aubin}}}]{Langlands92}%
  \BibitemOpen
  \bibfield  {author} {\bibinfo {author} {\bibfnamefont {R.~P.}\ \bibnamefont
  {Langlands}}, \bibinfo {author} {\bibfnamefont {C.}~\bibnamefont {Pichet}},
  \bibinfo {author} {\bibnamefont {{Ph. Pouliot}}}, \ and\ \bibinfo {author}
  {\bibfnamefont {Y.}~\bibnamefont {{Saint-Aubin}}},\ }\href {\doibase
  10.1007/BF01049720} {\bibfield  {journal} {\bibinfo  {journal} {J. Stat.
  Phys.}\ }\textbf {\bibinfo {volume} {67}},\ \bibinfo {pages} {553} (\bibinfo
  {year} {1992})}\BibitemShut {NoStop}%
\bibitem [{\citenamefont {Pinson}(1994)}]{Pinson94}%
  \BibitemOpen
  \bibfield  {author} {\bibinfo {author} {\bibfnamefont {H.~T.}\ \bibnamefont
  {Pinson}},\ }\href@noop {} {\bibfield  {journal} {\bibinfo  {journal} {J.
  Stat. Phys.}\ }\textbf {\bibinfo {volume} {75}},\ \bibinfo {pages} {1167}
  (\bibinfo {year} {1994})}\BibitemShut {NoStop}%
\bibitem [{\citenamefont {Watts}(1996)}]{Watts96}%
  \BibitemOpen
  \bibfield  {author} {\bibinfo {author} {\bibfnamefont {G.~M.~T.}\
  \bibnamefont {Watts}},\ }\href@noop {} {\bibfield  {journal} {\bibinfo
  {journal} {J. Phys. A}\ }\textbf {\bibinfo {volume} {29}},\ \bibinfo {pages}
  {L363} (\bibinfo {year} {1996})}\BibitemShut {NoStop}%
\bibitem [{\citenamefont {Simmons}\ \emph {et~al.}(2007)\citenamefont
  {Simmons}, \citenamefont {Kleban},\ and\ \citenamefont {Ziff}}]{Simmons07}%
  \BibitemOpen
  \bibfield  {author} {\bibinfo {author} {\bibfnamefont {J.~J.~H.}\
  \bibnamefont {Simmons}}, \bibinfo {author} {\bibfnamefont {P.}~\bibnamefont
  {Kleban}}, \ and\ \bibinfo {author} {\bibfnamefont {R.~M.}\ \bibnamefont
  {Ziff}},\ }\href@noop {} {\bibfield  {journal} {\bibinfo  {journal} {J. Phys.
  A}\ }\textbf {\bibinfo {volume} {40}},\ \bibinfo {pages} {F771} (\bibinfo
  {year} {2007})}\BibitemShut {NoStop}%
\bibitem [{\citenamefont {Schramm}(2000)}]{Schramm00}%
  \BibitemOpen
  \bibfield  {author} {\bibinfo {author} {\bibfnamefont {O.}~\bibnamefont
  {Schramm}},\ }\href {\doibase 10.1007/BF02803524} {\bibfield  {journal}
  {\bibinfo  {journal} {Israel J. Math.}\ }\textbf {\bibinfo {volume} {118}},\
  \bibinfo {pages} {221} (\bibinfo {year} {2000})}\BibitemShut {NoStop}%
\bibitem [{\citenamefont {Schramm}(2001)}]{Schramm01}%
  \BibitemOpen
  \bibfield  {author} {\bibinfo {author} {\bibfnamefont {O.}~\bibnamefont
  {Schramm}},\ }\href {\doibase 10.1214/ECP.v6-1041} {\bibfield  {journal}
  {\bibinfo  {journal} {Electron. Commun. Prob.}\ }\textbf {\bibinfo {volume}
  {6}},\ \bibinfo {pages} {115} (\bibinfo {year} {2001})}\BibitemShut {NoStop}%
\bibitem [{\citenamefont {Schramm}\ and\ \citenamefont
  {Sheffield}(2009)}]{Schramm09}%
  \BibitemOpen
  \bibfield  {author} {\bibinfo {author} {\bibfnamefont {O.}~\bibnamefont
  {Schramm}}\ and\ \bibinfo {author} {\bibfnamefont {S.}~\bibnamefont
  {Sheffield}},\ }\href {\doibase 10.1007/s11511-009-0034-y} {\bibfield
  {journal} {\bibinfo  {journal} {Acta Math.}\ }\textbf {\bibinfo {volume}
  {202}},\ \bibinfo {pages} {21} (\bibinfo {year} {2009})}\BibitemShut
  {NoStop}%
\bibitem [{\citenamefont {Smirnov}(2001)}]{Smirnov01}%
  \BibitemOpen
  \bibfield  {author} {\bibinfo {author} {\bibfnamefont {S.}~\bibnamefont
  {Smirnov}},\ }\href {\doibase 10.1016/S0764-4442(01)01991-7} {\bibfield
  {journal} {\bibinfo  {journal} {C. R. Acad. Sci. Paris I}\ }\textbf {\bibinfo
  {volume} {333}},\ \bibinfo {pages} {239} (\bibinfo {year}
  {2001})}\BibitemShut {NoStop}%
\bibitem [{\citenamefont {Smirnov}\ and\ \citenamefont
  {Werner}(2001)}]{Smirnov01b}%
  \BibitemOpen
  \bibfield  {author} {\bibinfo {author} {\bibfnamefont {S.}~\bibnamefont
  {Smirnov}}\ and\ \bibinfo {author} {\bibfnamefont {W.}~\bibnamefont
  {Werner}},\ }\href@noop {} {\bibfield  {journal} {\bibinfo  {journal} {Math.
  Res. Lett.}\ }\textbf {\bibinfo {volume} {8}},\ \bibinfo {pages} {729}
  (\bibinfo {year} {2001})}\BibitemShut {NoStop}%
\bibitem [{\citenamefont {Smirnov}(2006)}]{Smirnov06}%
  \BibitemOpen
  \bibfield  {author} {\bibinfo {author} {\bibfnamefont {S.}~\bibnamefont
  {Smirnov}},\ }in\ \href@noop {} {\emph {\bibinfo {booktitle} {Proceedings of
  the International Congress of Mathematicians, Madrid, Spain, 2006}}},\
  \bibinfo {editor} {edited by\ \bibinfo {editor} {\bibfnamefont
  {M.}~\bibnamefont {{Sanz-Sol\'e}}}, \bibinfo {editor} {\bibfnamefont
  {J.}~\bibnamefont {Soria}}, \bibinfo {editor} {\bibfnamefont {J.~L.}\
  \bibnamefont {Varona}}, \ and\ \bibinfo {editor} {\bibfnamefont
  {J.}~\bibnamefont {Verdera}}}\ (\bibinfo  {publisher} {European Mathematical
  Society},\ \bibinfo {address} {Z\"urich},\ \bibinfo {year} {2006})\ p.\
  \bibinfo {pages} {1421}\BibitemShut {NoStop}%
\bibitem [{\citenamefont {Erd{\H o}s}\ and\ \citenamefont
  {R{\'e}nyi}(1960)}]{Erdos60}%
  \BibitemOpen
  \bibfield  {author} {\bibinfo {author} {\bibfnamefont {P.}~\bibnamefont
  {Erd{\H o}s}}\ and\ \bibinfo {author} {\bibfnamefont {A.}~\bibnamefont
  {R{\'e}nyi}},\ }\href@noop {} {\bibfield  {journal} {\bibinfo  {journal}
  {Publ. Math. Inst. Hung. Acad. Sci.}\ }\textbf {\bibinfo {volume} {5}},\
  \bibinfo {pages} {17} (\bibinfo {year} {1960})}\BibitemShut {NoStop}%
\bibitem [{\citenamefont {Moore}\ and\ \citenamefont {Newman}(2000)}]{Moore00}%
  \BibitemOpen
  \bibfield  {author} {\bibinfo {author} {\bibfnamefont {C.}~\bibnamefont
  {Moore}}\ and\ \bibinfo {author} {\bibfnamefont {M.~E.~J.}\ \bibnamefont
  {Newman}},\ }\href@noop {} {\bibfield  {journal} {\bibinfo  {journal} {Phys.
  Rev. E}\ }\textbf {\bibinfo {volume} {62}},\ \bibinfo {pages} {7059}
  (\bibinfo {year} {2000})}\BibitemShut {NoStop}%
\bibitem [{\citenamefont {Achlioptas}\ \emph {et~al.}(2009)\citenamefont
  {Achlioptas}, \citenamefont {D'Souza},\ and\ \citenamefont
  {Spencer}}]{Achlioptas09}%
  \BibitemOpen
  \bibfield  {author} {\bibinfo {author} {\bibfnamefont {D.}~\bibnamefont
  {Achlioptas}}, \bibinfo {author} {\bibfnamefont {R.~M.}\ \bibnamefont
  {D'Souza}}, \ and\ \bibinfo {author} {\bibfnamefont {J.}~\bibnamefont
  {Spencer}},\ }\href {\doibase 10.1126/science.1167782} {\bibfield  {journal}
  {\bibinfo  {journal} {Science}\ }\textbf {\bibinfo {volume} {323}},\ \bibinfo
  {pages} {1453} (\bibinfo {year} {2009})}\BibitemShut {NoStop}%
\bibitem [{\citenamefont {Sykes}\ and\ \citenamefont {Essam}(1963)}]{Sykes63}%
  \BibitemOpen
  \bibfield  {author} {\bibinfo {author} {\bibfnamefont {M.~F.}\ \bibnamefont
  {Sykes}}\ and\ \bibinfo {author} {\bibfnamefont {J.~W.}\ \bibnamefont
  {Essam}},\ }\href {\doibase 10.1103/PhysRevLett.10.3} {\bibfield  {journal}
  {\bibinfo  {journal} {Phys. Rev. Lett.}\ }\textbf {\bibinfo {volume} {10}},\
  \bibinfo {pages} {3} (\bibinfo {year} {1963})}\BibitemShut {NoStop}%
\bibitem [{\citenamefont {Ziff}\ and\ \citenamefont {Scullard}(2006)}]{Ziff06}%
  \BibitemOpen
  \bibfield  {author} {\bibinfo {author} {\bibfnamefont {R.~M.}\ \bibnamefont
  {Ziff}}\ and\ \bibinfo {author} {\bibfnamefont {C.~R.}\ \bibnamefont
  {Scullard}},\ }\href {\doibase 10.1088/0305-4470/39/49/003} {\bibfield
  {journal} {\bibinfo  {journal} {J. Phys. A}\ }\textbf {\bibinfo {volume}
  {39}},\ \bibinfo {pages} {15083} (\bibinfo {year} {2006})}\BibitemShut
  {NoStop}%
\bibitem [{\citenamefont {Grimmett}\ and\ \citenamefont
  {Manolescu}(2013)}]{Grimmett13}%
  \BibitemOpen
  \bibfield  {author} {\bibinfo {author} {\bibfnamefont {G.~R.}\ \bibnamefont
  {Grimmett}}\ and\ \bibinfo {author} {\bibfnamefont {I.}~\bibnamefont
  {Manolescu}},\ }\href {\doibase 0.1007/s00440-013-0507-y} {\bibfield
  {journal} {\bibinfo  {journal} {Prob. Thoery Relat. Fields}\ } (\bibinfo
  {year} {2013}),\ 0.1007/s00440-013-0507-y},\ \bibinfo {note} {publ.
  online}\BibitemShut {NoStop}%
\bibitem [{\citenamefont {Ziff}\ \emph {et~al.}(2012)\citenamefont {Ziff},
  \citenamefont {Scullard}, \citenamefont {Wierman},\ and\ \citenamefont
  {Sedlock}}]{Ziff12}%
  \BibitemOpen
  \bibfield  {author} {\bibinfo {author} {\bibfnamefont {R.~M.}\ \bibnamefont
  {Ziff}}, \bibinfo {author} {\bibfnamefont {C.~R.}\ \bibnamefont {Scullard}},
  \bibinfo {author} {\bibfnamefont {J.~C.}\ \bibnamefont {Wierman}}, \ and\
  \bibinfo {author} {\bibfnamefont {M.~R.~A.}\ \bibnamefont {Sedlock}},\ }\href
  {\doibase 10.1088/1751-8113/45/49/494005} {\bibfield  {journal} {\bibinfo
  {journal} {J. Phys. A}\ }\textbf {\bibinfo {volume} {45}},\ \bibinfo {pages}
  {494005} (\bibinfo {year} {2012})}\BibitemShut {NoStop}%
\bibitem [{\citenamefont {Wu}(1979)}]{Wu79}%
  \BibitemOpen
  \bibfield  {author} {\bibinfo {author} {\bibfnamefont {F.~Y.}\ \bibnamefont
  {Wu}},\ }\href {\doibase 10.1088/0022-3719/12/17/002} {\bibfield  {journal}
  {\bibinfo  {journal} {J. Phys. C}\ }\textbf {\bibinfo {volume} {12}},\
  \bibinfo {pages} {L645} (\bibinfo {year} {1979})}\BibitemShut {NoStop}%
\bibitem [{\citenamefont {Jacobsen}\ and\ \citenamefont
  {Scullard}(2013)}]{Jacobsen13}%
  \BibitemOpen
  \bibfield  {author} {\bibinfo {author} {\bibfnamefont {J.~L.}\ \bibnamefont
  {Jacobsen}}\ and\ \bibinfo {author} {\bibfnamefont {C.~R.}\ \bibnamefont
  {Scullard}},\ }\href {\doibase 10.1088/1751-8113/46/7/075001} {\bibfield
  {journal} {\bibinfo  {journal} {J. Phys. A}\ }\textbf {\bibinfo {volume}
  {46}},\ \bibinfo {pages} {075001} (\bibinfo {year} {2013})}\BibitemShut
  {NoStop}%
\bibitem [{\citenamefont {Jacobsen}(2014)}]{Jacobsen14}%
  \BibitemOpen
  \bibfield  {author} {\bibinfo {author} {\bibfnamefont {J.~L.}\ \bibnamefont
  {Jacobsen}},\ }\href {\doibase 10.1088/1751-8113/47/13/135001} {\bibfield
  {journal} {\bibinfo  {journal} {J. Phys. A}\ }\textbf {\bibinfo {volume}
  {47}},\ \bibinfo {pages} {135001} (\bibinfo {year} {2014})}\BibitemShut
  {NoStop}%
\bibitem [{\citenamefont {Miyazima}(2005)}]{Miyazima06}%
  \BibitemOpen
  \bibfield  {author} {\bibinfo {author} {\bibfnamefont {S.}~\bibnamefont
  {Miyazima}},\ }\href@noop {} {\bibfield  {journal} {\bibinfo  {journal}
  {Bulletin, Institute of Science and Technology Research}\ }\textbf {\bibinfo
  {volume} {17}},\ \bibinfo {pages} {119} (\bibinfo {year} {2005})}\BibitemShut
  {NoStop}%
\bibitem [{\citenamefont {Grassberger}(1999{\natexlab{b}})}]{Grassberger99b}%
  \BibitemOpen
  \bibfield  {author} {\bibinfo {author} {\bibfnamefont {P.}~\bibnamefont
  {Grassberger}},\ }\href {\doibase 10.1088/0305-4470/32/35/301} {\bibfield
  {journal} {\bibinfo  {journal} {J. Phys. A}\ }\textbf {\bibinfo {volume}
  {32}},\ \bibinfo {pages} {6233} (\bibinfo {year}
  {1999}{\natexlab{b}})}\BibitemShut {NoStop}%
\bibitem [{\citenamefont {Ziff}(1999)}]{Ziff99b}%
  \BibitemOpen
  \bibfield  {author} {\bibinfo {author} {\bibfnamefont {R.~M.}\ \bibnamefont
  {Ziff}},\ }\href {\doibase 10.1088/0305-4470/32/43/101} {\bibfield  {journal}
  {\bibinfo  {journal} {J. Phys. A}\ }\textbf {\bibinfo {volume} {32}},\
  \bibinfo {pages} {L457} (\bibinfo {year} {1999})}\BibitemShut {NoStop}%
\bibitem [{\citenamefont {Ziff}()}]{Ziff11c}%
  \BibitemOpen
  \bibfield  {author} {\bibinfo {author} {\bibfnamefont {R.~M.}\ \bibnamefont
  {Ziff}},\ }\href@noop {} {}\Eprint {http://arxiv.org/abs/arXiv:1103.3243}
  {arXiv:1103.3243} \BibitemShut {NoStop}%
\bibitem [{\citenamefont {Aharony}\ and\ \citenamefont
  {Stauffer}(1997)}]{Aharony97}%
  \BibitemOpen
  \bibfield  {author} {\bibinfo {author} {\bibfnamefont {A.}~\bibnamefont
  {Aharony}}\ and\ \bibinfo {author} {\bibfnamefont {D.}~\bibnamefont
  {Stauffer}},\ }\href {\doibase 10.1088/0305-4470/30/10/001} {\bibfield
  {journal} {\bibinfo  {journal} {J. Phys. A}\ }\textbf {\bibinfo {volume}
  {30}},\ \bibinfo {pages} {L301} (\bibinfo {year} {1997})}\BibitemShut
  {NoStop}%
\bibitem [{\citenamefont {{de Souza}}\ \emph {et~al.}(2011)\citenamefont {{de
  Souza}}, \citenamefont {Tom\'e},\ and\ \citenamefont {Ziff}}]{deSouza11}%
  \BibitemOpen
  \bibfield  {author} {\bibinfo {author} {\bibfnamefont {D.~R.}\ \bibnamefont
  {{de Souza}}}, \bibinfo {author} {\bibfnamefont {T.}~\bibnamefont {Tom\'e}},
  \ and\ \bibinfo {author} {\bibfnamefont {R.~M.}\ \bibnamefont {Ziff}},\
  }\href {\doibase 10.1088/1742-5468/2011/03/P03006} {\bibfield  {journal}
  {\bibinfo  {journal} {J. Stat. Mech.}\ ,\ \bibinfo {pages} {P03006}}
  (\bibinfo {year} {2011})}\BibitemShut {NoStop}%
\bibitem [{\citenamefont {Privman}\ \emph {et~al.}(1991)\citenamefont
  {Privman}, \citenamefont {Hohenberg},\ and\ \citenamefont
  {Aharony}}]{Privman91}%
  \BibitemOpen
  \bibfield  {author} {\bibinfo {author} {\bibfnamefont {V.}~\bibnamefont
  {Privman}}, \bibinfo {author} {\bibfnamefont {P.~C.}\ \bibnamefont
  {Hohenberg}}, \ and\ \bibinfo {author} {\bibfnamefont {A.}~\bibnamefont
  {Aharony}},\ }in\ \href@noop {} {\emph {\bibinfo {booktitle} {Phase
  transitions and critical phenomena}}},\ Vol.~\bibinfo {volume} {14},\
  \bibinfo {editor} {edited by\ \bibinfo {editor} {\bibfnamefont
  {C.}~\bibnamefont {Domb}}\ and\ \bibinfo {editor} {\bibfnamefont {J.~L.}\
  \bibnamefont {Lebowitz}}}\ (\bibinfo  {publisher} {Academic},\ \bibinfo
  {address} {NY},\ \bibinfo {year} {1991})\BibitemShut {NoStop}%
\bibitem [{\citenamefont {Chayes}\ \emph {et~al.}(1989)\citenamefont {Chayes},
  \citenamefont {Chayes}, \citenamefont {Grimmett}, \citenamefont {Kesten},\
  and\ \citenamefont {Schonmann}}]{Chayes89}%
  \BibitemOpen
  \bibfield  {author} {\bibinfo {author} {\bibfnamefont {J.~T.}\ \bibnamefont
  {Chayes}}, \bibinfo {author} {\bibfnamefont {L.}~\bibnamefont {Chayes}},
  \bibinfo {author} {\bibfnamefont {G.~R.}\ \bibnamefont {Grimmett}}, \bibinfo
  {author} {\bibfnamefont {H.}~\bibnamefont {Kesten}}, \ and\ \bibinfo {author}
  {\bibfnamefont {R.~H.}\ \bibnamefont {Schonmann}},\ }\href {\doibase
  10.1214/aop/1176991155} {\bibfield  {journal} {\bibinfo  {journal} {Ann.
  Probab.}\ }\textbf {\bibinfo {volume} {17}},\ \bibinfo {pages} {1277}
  (\bibinfo {year} {1989})}\BibitemShut {NoStop}%
\bibitem [{\citenamefont {Seaton}(2001)}]{Seaton01}%
  \BibitemOpen
  \bibfield  {author} {\bibinfo {author} {\bibfnamefont {K.~A.}\ \bibnamefont
  {Seaton}},\ }\href {\doibase 10.1088/0305-4470/34/50/105} {\bibfield
  {journal} {\bibinfo  {journal} {J. Phys. A}\ }\textbf {\bibinfo {volume}
  {34}},\ \bibinfo {pages} {L759} (\bibinfo {year} {2001})}\BibitemShut
  {NoStop}%
\bibitem [{\citenamefont {Delfino}\ and\ \citenamefont
  {Cardy}(1998)}]{Delfino98}%
  \BibitemOpen
  \bibfield  {author} {\bibinfo {author} {\bibfnamefont {G.}~\bibnamefont
  {Delfino}}\ and\ \bibinfo {author} {\bibfnamefont {J.~L.}\ \bibnamefont
  {Cardy}},\ }\href {\doibase 10.1016/S0550-3213(98)00144-8} {\bibfield
  {journal} {\bibinfo  {journal} {Nucl. Phys. B}\ }\textbf {\bibinfo {volume}
  {519}},\ \bibinfo {pages} {551} (\bibinfo {year} {1998})}\BibitemShut
  {NoStop}%
\bibitem [{\citenamefont {Cardy}\ and\ \citenamefont {Ziff}(2003)}]{Cardy03}%
  \BibitemOpen
  \bibfield  {author} {\bibinfo {author} {\bibfnamefont {J.}~\bibnamefont
  {Cardy}}\ and\ \bibinfo {author} {\bibfnamefont {R.~M.}\ \bibnamefont
  {Ziff}},\ }\href {\doibase 10.1023/A:1021069209656} {\bibfield  {journal}
  {\bibinfo  {journal} {J. Stat. Phys.}\ }\textbf {\bibinfo {volume} {110}},\
  \bibinfo {pages} {1} (\bibinfo {year} {2003})}\BibitemShut {NoStop}%
\bibitem [{\citenamefont {Ziff}\ \emph {et~al.}(1999)\citenamefont {Ziff},
  \citenamefont {Lorenz},\ and\ \citenamefont {Kleban}}]{Ziff99c}%
  \BibitemOpen
  \bibfield  {author} {\bibinfo {author} {\bibfnamefont {R.~M.}\ \bibnamefont
  {Ziff}}, \bibinfo {author} {\bibfnamefont {C.~D.}\ \bibnamefont {Lorenz}}, \
  and\ \bibinfo {author} {\bibfnamefont {P.}~\bibnamefont {Kleban}},\ }\href
  {\doibase 10.1016/S0378-4371(98)00569-X} {\bibfield  {journal} {\bibinfo
  {journal} {Physica A}\ }\textbf {\bibinfo {volume} {266}},\ \bibinfo {pages}
  {17} (\bibinfo {year} {1999})}\BibitemShut {NoStop}%
\bibitem [{\citenamefont {Temperley}\ and\ \citenamefont
  {Lieb}(1971)}]{Temperley71}%
  \BibitemOpen
  \bibfield  {author} {\bibinfo {author} {\bibfnamefont {H.~N.~V.}\
  \bibnamefont {Temperley}}\ and\ \bibinfo {author} {\bibfnamefont {E.~H.}\
  \bibnamefont {Lieb}},\ }\href {\doibase 10.1098/rspa.1971.0067} {\bibfield
  {journal} {\bibinfo  {journal} {Proc. R. Soc. Lond. A}\ }\textbf {\bibinfo
  {volume} {322}},\ \bibinfo {pages} {251} (\bibinfo {year}
  {1971})}\BibitemShut {NoStop}%
\bibitem [{\citenamefont {Ziff}\ \emph {et~al.}(1997)\citenamefont {Ziff},
  \citenamefont {Finch},\ and\ \citenamefont {Adamchik}}]{Ziff97b}%
  \BibitemOpen
  \bibfield  {author} {\bibinfo {author} {\bibfnamefont {R.~M.}\ \bibnamefont
  {Ziff}}, \bibinfo {author} {\bibfnamefont {S.~R.}\ \bibnamefont {Finch}}, \
  and\ \bibinfo {author} {\bibfnamefont {V.~S.}\ \bibnamefont {Adamchik}},\
  }\href {\doibase 10.1103/PhysRevLett.79.3447} {\bibfield  {journal} {\bibinfo
   {journal} {Phys. Rev. Lett.}\ }\textbf {\bibinfo {volume} {79}},\ \bibinfo
  {pages} {3447} (\bibinfo {year} {1997})}\BibitemShut {NoStop}%
\bibitem [{\citenamefont {Baxter}\ \emph {et~al.}(1978)\citenamefont {Baxter},
  \citenamefont {Temperley},\ and\ \citenamefont {Ashley}}]{Baxter78}%
  \BibitemOpen
  \bibfield  {author} {\bibinfo {author} {\bibfnamefont {R.~J.}\ \bibnamefont
  {Baxter}}, \bibinfo {author} {\bibfnamefont {H.~N.~V.}\ \bibnamefont
  {Temperley}}, \ and\ \bibinfo {author} {\bibfnamefont {S.~E.}\ \bibnamefont
  {Ashley}},\ }\href {\doibase 10.1098/rspa.1978.0026} {\bibfield  {journal}
  {\bibinfo  {journal} {Proc. R. Soc. Lond. A}\ }\textbf {\bibinfo {volume}
  {358}},\ \bibinfo {pages} {535} (\bibinfo {year} {1978})}\BibitemShut
  {NoStop}%
\bibitem [{\citenamefont {Isichenko}(1992)}]{Isichenko92}%
  \BibitemOpen
  \bibfield  {author} {\bibinfo {author} {\bibfnamefont {M.~B.}\ \bibnamefont
  {Isichenko}},\ }\href {\doibase 10.1103/RevModPhys.64.961} {\bibfield
  {journal} {\bibinfo  {journal} {Rev. Mod. Phys.}\ }\textbf {\bibinfo {volume}
  {64}},\ \bibinfo {pages} {961} (\bibinfo {year} {1992})}\BibitemShut
  {NoStop}%
\bibitem [{\citenamefont {Sahimi}(1993)}]{Sahimi93}%
  \BibitemOpen
  \bibfield  {author} {\bibinfo {author} {\bibfnamefont {M.}~\bibnamefont
  {Sahimi}},\ }\href {\doibase 10.1103/RevModPhys.65.1393} {\bibfield
  {journal} {\bibinfo  {journal} {Rev. Mod. Phys.}\ }\textbf {\bibinfo {volume}
  {65}},\ \bibinfo {pages} {1393} (\bibinfo {year} {1993})}\BibitemShut
  {NoStop}%
\bibitem [{\citenamefont {Weinrib}(1984)}]{Weinrib84}%
  \BibitemOpen
  \bibfield  {author} {\bibinfo {author} {\bibfnamefont {A.}~\bibnamefont
  {Weinrib}},\ }\href {\doibase 10.1103/PhysRevB.29.387} {\bibfield  {journal}
  {\bibinfo  {journal} {Phys. Rev. B}\ }\textbf {\bibinfo {volume} {29}},\
  \bibinfo {pages} {387} (\bibinfo {year} {1984})}\BibitemShut {NoStop}%
\bibitem [{\citenamefont {Prakash}\ \emph {et~al.}(1992)\citenamefont
  {Prakash}, \citenamefont {Havlin}, \citenamefont {Schwartz},\ and\
  \citenamefont {Stanley}}]{Prakash92}%
  \BibitemOpen
  \bibfield  {author} {\bibinfo {author} {\bibfnamefont {S.}~\bibnamefont
  {Prakash}}, \bibinfo {author} {\bibfnamefont {S.}~\bibnamefont {Havlin}},
  \bibinfo {author} {\bibfnamefont {M.}~\bibnamefont {Schwartz}}, \ and\
  \bibinfo {author} {\bibfnamefont {H.~E.}\ \bibnamefont {Stanley}},\ }\href
  {\doibase 10.1103/PhysRevA.46.R1724} {\bibfield  {journal} {\bibinfo
  {journal} {Phys. Rev. A}\ }\textbf {\bibinfo {volume} {46}},\ \bibinfo
  {pages} {R1724} (\bibinfo {year} {1992})}\BibitemShut {NoStop}%
\bibitem [{\citenamefont {Schmittbuhl}\ \emph {et~al.}(1993)\citenamefont
  {Schmittbuhl}, \citenamefont {Vilotte},\ and\ \citenamefont
  {Roux}}]{Schmittbuhl93}%
  \BibitemOpen
  \bibfield  {author} {\bibinfo {author} {\bibfnamefont {J.}~\bibnamefont
  {Schmittbuhl}}, \bibinfo {author} {\bibfnamefont {J.-P.}\ \bibnamefont
  {Vilotte}}, \ and\ \bibinfo {author} {\bibfnamefont {S.}~\bibnamefont
  {Roux}},\ }\href {\doibase 10.1088/0305-4470/26/22/014} {\bibfield  {journal}
  {\bibinfo  {journal} {J. Phys. A}\ }\textbf {\bibinfo {volume} {26}},\
  \bibinfo {pages} {6115} (\bibinfo {year} {1993})}\BibitemShut {NoStop}%
\bibitem [{\citenamefont {Sahimi}(1994{\natexlab{b}})}]{Sahimi94a}%
  \BibitemOpen
  \bibfield  {author} {\bibinfo {author} {\bibfnamefont {M.}~\bibnamefont
  {Sahimi}},\ }\href {\doibase 10.1051/jp1:1994107} {\bibfield  {journal}
  {\bibinfo  {journal} {J. Phys. I France}\ }\textbf {\bibinfo {volume} {4}},\
  \bibinfo {pages} {1263} (\bibinfo {year} {1994}{\natexlab{b}})}\BibitemShut
  {NoStop}%
\bibitem [{\citenamefont {Du}\ \emph {et~al.}(1996)\citenamefont {Du},
  \citenamefont {Satik},\ and\ \citenamefont {Yortsos}}]{Du04}%
  \BibitemOpen
  \bibfield  {author} {\bibinfo {author} {\bibfnamefont {C.}~\bibnamefont
  {Du}}, \bibinfo {author} {\bibfnamefont {C.}~\bibnamefont {Satik}}, \ and\
  \bibinfo {author} {\bibfnamefont {Y.~C.}\ \bibnamefont {Yortsos}},\ }\href
  {\doibase 10.1002/aic.690420831} {\bibfield  {journal} {\bibinfo  {journal}
  {AIChE Journal}\ }\textbf {\bibinfo {volume} {42}},\ \bibinfo {pages} {2392}
  (\bibinfo {year} {1996})}\BibitemShut {NoStop}%
\bibitem [{\citenamefont {Makse}\ \emph {et~al.}(1995)\citenamefont {Makse},
  \citenamefont {Havlin},\ and\ \citenamefont {Stanley}}]{Makse95}%
  \BibitemOpen
  \bibfield  {author} {\bibinfo {author} {\bibfnamefont {H.~A.}\ \bibnamefont
  {Makse}}, \bibinfo {author} {\bibfnamefont {S.}~\bibnamefont {Havlin}}, \
  and\ \bibinfo {author} {\bibfnamefont {H.~E.}\ \bibnamefont {Stanley}},\
  }\href {\doibase 10.1038/377608a0} {\bibfield  {journal} {\bibinfo  {journal}
  {Nature}\ }\textbf {\bibinfo {volume} {377}},\ \bibinfo {pages} {608}
  (\bibinfo {year} {1995})}\BibitemShut {NoStop}%
\bibitem [{\citenamefont {Makse}\ \emph {et~al.}(1998)\citenamefont {Makse},
  \citenamefont {{Andrade Jr.}}, \citenamefont {Batty}, \citenamefont
  {Havlin},\ and\ \citenamefont {Stanley}}]{Makse98}%
  \BibitemOpen
  \bibfield  {author} {\bibinfo {author} {\bibfnamefont {H.~A.}\ \bibnamefont
  {Makse}}, \bibinfo {author} {\bibfnamefont {J.~S.}\ \bibnamefont {{Andrade
  Jr.}}}, \bibinfo {author} {\bibfnamefont {M.}~\bibnamefont {Batty}}, \bibinfo
  {author} {\bibfnamefont {S.}~\bibnamefont {Havlin}}, \ and\ \bibinfo {author}
  {\bibfnamefont {H.~E.}\ \bibnamefont {Stanley}},\ }\href {\doibase
  10.1103/PhysRevE.58.7054} {\bibfield  {journal} {\bibinfo  {journal} {Phys.
  Rev. E}\ }\textbf {\bibinfo {volume} {58}},\ \bibinfo {pages} {7054}
  (\bibinfo {year} {1998})}\BibitemShut {NoStop}%
\bibitem [{\citenamefont {Sahimi}(1998)}]{Sahimi98}%
  \BibitemOpen
  \bibfield  {author} {\bibinfo {author} {\bibfnamefont {M.}~\bibnamefont
  {Sahimi}},\ }\href {\doibase 10.1016/S0370-1573(98)00024-6} {\bibfield
  {journal} {\bibinfo  {journal} {Phys. Rep.}\ }\textbf {\bibinfo {volume}
  {306}},\ \bibinfo {pages} {213} (\bibinfo {year} {1998})}\BibitemShut
  {NoStop}%
\bibitem [{\citenamefont {Makse}\ \emph {et~al.}(2000)\citenamefont {Makse},
  \citenamefont {{Andrade Jr.}},\ and\ \citenamefont {Stanley}}]{Makse00}%
  \BibitemOpen
  \bibfield  {author} {\bibinfo {author} {\bibfnamefont {H.~A.}\ \bibnamefont
  {Makse}}, \bibinfo {author} {\bibfnamefont {J.~S.}\ \bibnamefont {{Andrade
  Jr.}}}, \ and\ \bibinfo {author} {\bibfnamefont {H.~E.}\ \bibnamefont
  {Stanley}},\ }\href {\doibase 10.1103/PhysRevE.61.583} {\bibfield  {journal}
  {\bibinfo  {journal} {Phys. Rev. E}\ }\textbf {\bibinfo {volume} {61}},\
  \bibinfo {pages} {583} (\bibinfo {year} {2000})}\BibitemShut {NoStop}%
\bibitem [{\citenamefont {Ara\'ujo}\ \emph {et~al.}(2002)\citenamefont
  {Ara\'ujo}, \citenamefont {Moreira}, \citenamefont {Makse}, \citenamefont
  {Stanley},\ and\ \citenamefont {{Andrade Jr.}}}]{Araujo02}%
  \BibitemOpen
  \bibfield  {author} {\bibinfo {author} {\bibfnamefont {A.~D.}\ \bibnamefont
  {Ara\'ujo}}, \bibinfo {author} {\bibfnamefont {A.~A.}\ \bibnamefont
  {Moreira}}, \bibinfo {author} {\bibfnamefont {H.~A.}\ \bibnamefont {Makse}},
  \bibinfo {author} {\bibfnamefont {H.~E.}\ \bibnamefont {Stanley}}, \ and\
  \bibinfo {author} {\bibfnamefont {J.~S.}\ \bibnamefont {{Andrade Jr.}}},\
  }\href {\doibase 10.1103/PhysRevE.66.046304} {\bibfield  {journal} {\bibinfo
  {journal} {Phys. Rev. E}\ }\textbf {\bibinfo {volume} {66}},\ \bibinfo
  {pages} {046304} (\bibinfo {year} {2002})}\BibitemShut {NoStop}%
\bibitem [{\citenamefont {Ara\'ujo}\ \emph {et~al.}(2003)\citenamefont
  {Ara\'ujo}, \citenamefont {Moreira}, \citenamefont {{Costa Filho}},\ and\
  \citenamefont {{Andrade Jr.}}}]{Araujo03}%
  \BibitemOpen
  \bibfield  {author} {\bibinfo {author} {\bibfnamefont {A.~D.}\ \bibnamefont
  {Ara\'ujo}}, \bibinfo {author} {\bibfnamefont {A.~A.}\ \bibnamefont
  {Moreira}}, \bibinfo {author} {\bibfnamefont {R.~N.}\ \bibnamefont {{Costa
  Filho}}}, \ and\ \bibinfo {author} {\bibfnamefont {J.~S.}\ \bibnamefont
  {{Andrade Jr.}}},\ }\href {\doibase 10.1103/PhysRevE.67.027102} {\bibfield
  {journal} {\bibinfo  {journal} {Phys. Rev. E}\ }\textbf {\bibinfo {volume}
  {67}},\ \bibinfo {pages} {027102} (\bibinfo {year} {2003})}\BibitemShut
  {NoStop}%
\bibitem [{\citenamefont {Sandler}\ \emph {et~al.}(2004)\citenamefont
  {Sandler}, \citenamefont {Maei},\ and\ \citenamefont {Kondev}}]{Sandler04}%
  \BibitemOpen
  \bibfield  {author} {\bibinfo {author} {\bibfnamefont {N.}~\bibnamefont
  {Sandler}}, \bibinfo {author} {\bibfnamefont {H.~R.}\ \bibnamefont {Maei}}, \
  and\ \bibinfo {author} {\bibfnamefont {J.}~\bibnamefont {Kondev}},\ }\href
  {\doibase 10.1103/PhysRevB.70.045309} {\bibfield  {journal} {\bibinfo
  {journal} {Phys. Rev. B}\ }\textbf {\bibinfo {volume} {70}},\ \bibinfo
  {pages} {045309} (\bibinfo {year} {2004})}\BibitemShut {NoStop}%
\bibitem [{\citenamefont {Schrenk}\ \emph
  {et~al.}(2013{\natexlab{b}})\citenamefont {Schrenk}, \citenamefont {Pos\'e},
  \citenamefont {Kranz}, \citenamefont {{van Kessenich}}, \citenamefont
  {Ara\'ujo},\ and\ \citenamefont {Herrmann}}]{Schrenk13b}%
  \BibitemOpen
  \bibfield  {author} {\bibinfo {author} {\bibfnamefont {K.~J.}\ \bibnamefont
  {Schrenk}}, \bibinfo {author} {\bibfnamefont {N.}~\bibnamefont {Pos\'e}},
  \bibinfo {author} {\bibfnamefont {J.~J.}\ \bibnamefont {Kranz}}, \bibinfo
  {author} {\bibfnamefont {L.~V.~M.}\ \bibnamefont {{van Kessenich}}}, \bibinfo
  {author} {\bibfnamefont {N.~A.~M.}\ \bibnamefont {Ara\'ujo}}, \ and\ \bibinfo
  {author} {\bibfnamefont {H.~J.}\ \bibnamefont {Herrmann}},\ }\href {\doibase
  10.1103/PhysRevE.88.052102} {\bibfield  {journal} {\bibinfo  {journal} {Phys.
  Rev. E}\ }\textbf {\bibinfo {volume} {88}},\ \bibinfo {pages} {052102}
  (\bibinfo {year} {2013}{\natexlab{b}})}\BibitemShut {NoStop}%
\bibitem [{\citenamefont {Auradou}\ \emph {et~al.}(1999)\citenamefont
  {Auradou}, \citenamefont {M{\aa}l{\o}y}, \citenamefont {Schmittbuhl},
  \citenamefont {Hansen},\ and\ \citenamefont {Bideau}}]{Auradou99}%
  \BibitemOpen
  \bibfield  {author} {\bibinfo {author} {\bibfnamefont {H.}~\bibnamefont
  {Auradou}}, \bibinfo {author} {\bibfnamefont {K.~J.}\ \bibnamefont
  {M{\aa}l{\o}y}}, \bibinfo {author} {\bibfnamefont {J.}~\bibnamefont
  {Schmittbuhl}}, \bibinfo {author} {\bibfnamefont {A.}~\bibnamefont {Hansen}},
  \ and\ \bibinfo {author} {\bibfnamefont {D.}~\bibnamefont {Bideau}},\ }\href
  {\doibase 10.1103/PhysRevE.60.7224} {\bibfield  {journal} {\bibinfo
  {journal} {Phys. Rev. E}\ }\textbf {\bibinfo {volume} {60}},\ \bibinfo
  {pages} {7224} (\bibinfo {year} {1999})}\BibitemShut {NoStop}%
\bibitem [{\citenamefont {Schmittbuhl}\ \emph {et~al.}(2000)\citenamefont
  {Schmittbuhl}, \citenamefont {Hansen}, \citenamefont {Auradou},\ and\
  \citenamefont {M{\aa}l{\o}y}}]{Schmittbuhl00}%
  \BibitemOpen
  \bibfield  {author} {\bibinfo {author} {\bibfnamefont {J.}~\bibnamefont
  {Schmittbuhl}}, \bibinfo {author} {\bibfnamefont {A.}~\bibnamefont {Hansen}},
  \bibinfo {author} {\bibfnamefont {H.}~\bibnamefont {Auradou}}, \ and\
  \bibinfo {author} {\bibfnamefont {K.~J.}\ \bibnamefont {M{\aa}l{\o}y}},\
  }\href {\doibase 10.1103/PhysRevE.61.3985} {\bibfield  {journal} {\bibinfo
  {journal} {Phys. Rev. E}\ }\textbf {\bibinfo {volume} {61}},\ \bibinfo
  {pages} {3985} (\bibinfo {year} {2000})}\BibitemShut {NoStop}%
\bibitem [{\citenamefont {Kalda}(2008)}]{Kalda08}%
  \BibitemOpen
  \bibfield  {author} {\bibinfo {author} {\bibfnamefont {J.}~\bibnamefont
  {Kalda}},\ }\href {\doibase 10.1209/0295-5075/84/46003} {\bibfield  {journal}
  {\bibinfo  {journal} {EPL}\ }\textbf {\bibinfo {volume} {84}},\ \bibinfo
  {pages} {46003} (\bibinfo {year} {2008})}\BibitemShut {NoStop}%
\bibitem [{\citenamefont {Kondev}\ \emph {et~al.}(2000)\citenamefont {Kondev},
  \citenamefont {Henley},\ and\ \citenamefont {Salinas}}]{Kondev00}%
  \BibitemOpen
  \bibfield  {author} {\bibinfo {author} {\bibfnamefont {J.}~\bibnamefont
  {Kondev}}, \bibinfo {author} {\bibfnamefont {C.~L.}\ \bibnamefont {Henley}},
  \ and\ \bibinfo {author} {\bibfnamefont {D.~G.}\ \bibnamefont {Salinas}},\
  }\href {\doibase 10.1103/PhysRevE.61.104} {\bibfield  {journal} {\bibinfo
  {journal} {Phys. Rev. E}\ }\textbf {\bibinfo {volume} {61}},\ \bibinfo
  {pages} {104} (\bibinfo {year} {2000})}\BibitemShut {NoStop}%
\bibitem [{\citenamefont {Schrenk}\ \emph
  {et~al.}(2012{\natexlab{a}})\citenamefont {Schrenk}, \citenamefont
  {Ara\'ujo}, \citenamefont {{Andrade Jr.}},\ and\ \citenamefont
  {Herrmann}}]{Schrenk12}%
  \BibitemOpen
  \bibfield  {author} {\bibinfo {author} {\bibfnamefont {K.~J.}\ \bibnamefont
  {Schrenk}}, \bibinfo {author} {\bibfnamefont {N.~A.~M.}\ \bibnamefont
  {Ara\'ujo}}, \bibinfo {author} {\bibfnamefont {J.~S.}\ \bibnamefont {{Andrade
  Jr.}}}, \ and\ \bibinfo {author} {\bibfnamefont {H.~J.}\ \bibnamefont
  {Herrmann}},\ }\href {\doibase 10.1038/srep00348} {\bibfield  {journal}
  {\bibinfo  {journal} {Sci. Rep.}\ }\textbf {\bibinfo {volume} {2}},\ \bibinfo
  {pages} {348} (\bibinfo {year} {2012}{\natexlab{a}})}\BibitemShut {NoStop}%
\bibitem [{\citenamefont {Newman}\ and\ \citenamefont {Ziff}(2001)}]{Newman01}%
  \BibitemOpen
  \bibfield  {author} {\bibinfo {author} {\bibfnamefont {M.~E.~J.}\
  \bibnamefont {Newman}}\ and\ \bibinfo {author} {\bibfnamefont {R.~M.}\
  \bibnamefont {Ziff}},\ }\href {\doibase 10.1103/PhysRevE.64.016706}
  {\bibfield  {journal} {\bibinfo  {journal} {Phys. Rev. E}\ }\textbf {\bibinfo
  {volume} {64}},\ \bibinfo {pages} {016706} (\bibinfo {year}
  {2001})}\BibitemShut {NoStop}%
\bibitem [{\citenamefont {Hu}\ \emph {et~al.}(2012)\citenamefont {Hu},
  \citenamefont {Bl{\"o}te},\ and\ \citenamefont {Deng}}]{Hu12}%
  \BibitemOpen
  \bibfield  {author} {\bibinfo {author} {\bibfnamefont {H.}~\bibnamefont
  {Hu}}, \bibinfo {author} {\bibfnamefont {H.~W.~J.}\ \bibnamefont
  {Bl{\"o}te}}, \ and\ \bibinfo {author} {\bibfnamefont {Y.}~\bibnamefont
  {Deng}},\ }\href {\doibase 10.1088/1751-8113/45/49/494006} {\bibfield
  {journal} {\bibinfo  {journal} {J. Phys. A}\ }\textbf {\bibinfo {volume}
  {45}},\ \bibinfo {pages} {494006} (\bibinfo {year} {2012})}\BibitemShut
  {NoStop}%
\bibitem [{\citenamefont {Peitgen}\ and\ \citenamefont
  {Saupe}(1988)}]{Peitgen88}%
  \BibitemOpen
  \bibinfo {editor} {\bibfnamefont {H.-O.}\ \bibnamefont {Peitgen}}\ and\
  \bibinfo {editor} {\bibfnamefont {D.}~\bibnamefont {Saupe}},\ eds.,\
  \href@noop {} {\emph {\bibinfo {title} {The Science of Fractal Images}}}\
  (\bibinfo  {publisher} {Springer},\ \bibinfo {address} {New York},\ \bibinfo
  {year} {1988})\BibitemShut {NoStop}%
\bibitem [{\citenamefont {Lauritsen}\ \emph {et~al.}(1993)\citenamefont
  {Lauritsen}, \citenamefont {Sahimi},\ and\ \citenamefont
  {Herrmann}}]{Lauritsen93}%
  \BibitemOpen
  \bibfield  {author} {\bibinfo {author} {\bibfnamefont {K.~B.}\ \bibnamefont
  {Lauritsen}}, \bibinfo {author} {\bibfnamefont {M.}~\bibnamefont {Sahimi}}, \
  and\ \bibinfo {author} {\bibfnamefont {H.~J.}\ \bibnamefont {Herrmann}},\
  }\href {\doibase 10.1103/PhysRevE.48.1272} {\bibfield  {journal} {\bibinfo
  {journal} {Phys. Rev. E}\ }\textbf {\bibinfo {volume} {48}},\ \bibinfo
  {pages} {1272} (\bibinfo {year} {1993})}\BibitemShut {NoStop}%
\bibitem [{\citenamefont {Makse}\ \emph {et~al.}(1996)\citenamefont {Makse},
  \citenamefont {Havlin}, \citenamefont {Schwartz},\ and\ \citenamefont
  {Stanley}}]{Makse96}%
  \BibitemOpen
  \bibfield  {author} {\bibinfo {author} {\bibfnamefont {H.~A.}\ \bibnamefont
  {Makse}}, \bibinfo {author} {\bibfnamefont {S.}~\bibnamefont {Havlin}},
  \bibinfo {author} {\bibfnamefont {M.}~\bibnamefont {Schwartz}}, \ and\
  \bibinfo {author} {\bibfnamefont {H.~E.}\ \bibnamefont {Stanley}},\ }\href
  {\doibase 10.1103/PhysRevE.53.5445} {\bibfield  {journal} {\bibinfo
  {journal} {Phys. Rev. E}\ }\textbf {\bibinfo {volume} {53}},\ \bibinfo
  {pages} {5445} (\bibinfo {year} {1996})}\BibitemShut {NoStop}%
\bibitem [{\citenamefont {Ballesteros}\ and\ \citenamefont
  {Parisi}(1999)}]{Ballesteros99b}%
  \BibitemOpen
  \bibfield  {author} {\bibinfo {author} {\bibfnamefont {H.~G.}\ \bibnamefont
  {Ballesteros}}\ and\ \bibinfo {author} {\bibfnamefont {G.}~\bibnamefont
  {Parisi}},\ }\href {\doibase 10.1103/PhysRevB.60.12912} {\bibfield  {journal}
  {\bibinfo  {journal} {Phys. Rev. B}\ }\textbf {\bibinfo {volume} {60}},\
  \bibinfo {pages} {12912} (\bibinfo {year} {1999})}\BibitemShut {NoStop}%
\bibitem [{\citenamefont {Malamud}\ and\ \citenamefont
  {Turcotte}(1999)}]{Malamud99}%
  \BibitemOpen
  \bibfield  {author} {\bibinfo {author} {\bibfnamefont {B.~D.}\ \bibnamefont
  {Malamud}}\ and\ \bibinfo {author} {\bibfnamefont {D.~L.}\ \bibnamefont
  {Turcotte}},\ }\href {\doibase 10.1016/S0378-3758(98)00249-3} {\bibfield
  {journal} {\bibinfo  {journal} {J. Stat. Plan. Infer.}\ }\textbf {\bibinfo
  {volume} {80}},\ \bibinfo {pages} {173} (\bibinfo {year} {1999})}\BibitemShut
  {NoStop}%
\bibitem [{\citenamefont {Oliveira}\ \emph {et~al.}(2011)\citenamefont
  {Oliveira}, \citenamefont {Schrenk}, \citenamefont {Ara\'ujo}, \citenamefont
  {Herrmann},\ and\ \citenamefont {{Andrade Jr.}}}]{Oliveira11}%
  \BibitemOpen
  \bibfield  {author} {\bibinfo {author} {\bibfnamefont {E.~A.}\ \bibnamefont
  {Oliveira}}, \bibinfo {author} {\bibfnamefont {K.~J.}\ \bibnamefont
  {Schrenk}}, \bibinfo {author} {\bibfnamefont {N.~A.~M.}\ \bibnamefont
  {Ara\'ujo}}, \bibinfo {author} {\bibfnamefont {H.~J.}\ \bibnamefont
  {Herrmann}}, \ and\ \bibinfo {author} {\bibfnamefont {J.~S.}\ \bibnamefont
  {{Andrade Jr.}}},\ }\href {\doibase 10.1103/PhysRevE.83.046113} {\bibfield
  {journal} {\bibinfo  {journal} {Phys. Rev. E}\ }\textbf {\bibinfo {volume}
  {83}},\ \bibinfo {pages} {046113} (\bibinfo {year} {2011})}\BibitemShut
  {NoStop}%
\bibitem [{\citenamefont {Ahrens}\ and\ \citenamefont
  {Hartmann}(2011)}]{Ahrens11}%
  \BibitemOpen
  \bibfield  {author} {\bibinfo {author} {\bibfnamefont {B.}~\bibnamefont
  {Ahrens}}\ and\ \bibinfo {author} {\bibfnamefont {A.~K.}\ \bibnamefont
  {Hartmann}},\ }\href {\doibase 10.1103/PhysRevB.84.144202} {\bibfield
  {journal} {\bibinfo  {journal} {Phys. Rev. B}\ }\textbf {\bibinfo {volume}
  {84}},\ \bibinfo {pages} {144202} (\bibinfo {year} {2011})}\BibitemShut
  {NoStop}%
\bibitem [{\citenamefont {Morais}\ \emph {et~al.}(2011)\citenamefont {Morais},
  \citenamefont {Oliveira}, \citenamefont {Ara\'ujo}, \citenamefont
  {Herrmann},\ and\ \citenamefont {{Andrade Jr.}}}]{Morais11}%
  \BibitemOpen
  \bibfield  {author} {\bibinfo {author} {\bibfnamefont {P.~A.}\ \bibnamefont
  {Morais}}, \bibinfo {author} {\bibfnamefont {E.~A.}\ \bibnamefont
  {Oliveira}}, \bibinfo {author} {\bibfnamefont {N.~A.~M.}\ \bibnamefont
  {Ara\'ujo}}, \bibinfo {author} {\bibfnamefont {H.~J.}\ \bibnamefont
  {Herrmann}}, \ and\ \bibinfo {author} {\bibfnamefont {J.~S.}\ \bibnamefont
  {{Andrade Jr.}}},\ }\href {\doibase 10.1103/PhysRevE.84.016102} {\bibfield
  {journal} {\bibinfo  {journal} {Phys. Rev. E}\ }\textbf {\bibinfo {volume}
  {84}},\ \bibinfo {pages} {016102} (\bibinfo {year} {2011})}\BibitemShut
  {NoStop}%
\bibitem [{\citenamefont {Weinrib}\ and\ \citenamefont
  {Halperin}(1983)}]{Weinrib83}%
  \BibitemOpen
  \bibfield  {author} {\bibinfo {author} {\bibfnamefont {A.}~\bibnamefont
  {Weinrib}}\ and\ \bibinfo {author} {\bibfnamefont {B.~I.}\ \bibnamefont
  {Halperin}},\ }\href {\doibase 10.1103/PhysRevB.27.413} {\bibfield  {journal}
  {\bibinfo  {journal} {Phys. Rev. B}\ }\textbf {\bibinfo {volume} {27}},\
  \bibinfo {pages} {413} (\bibinfo {year} {1983})}\BibitemShut {NoStop}%
\bibitem [{\citenamefont {Janke}\ and\ \citenamefont
  {Weigel}(2004)}]{Janke04b}%
  \BibitemOpen
  \bibfield  {author} {\bibinfo {author} {\bibfnamefont {W.}~\bibnamefont
  {Janke}}\ and\ \bibinfo {author} {\bibfnamefont {M.}~\bibnamefont {Weigel}},\
  }\href {\doibase 10.1103/PhysRevB.69.144208} {\bibfield  {journal} {\bibinfo
  {journal} {Phys. Rev. B}\ }\textbf {\bibinfo {volume} {69}},\ \bibinfo
  {pages} {144208} (\bibinfo {year} {2004})}\BibitemShut {NoStop}%
\bibitem [{\citenamefont {Mandre}\ and\ \citenamefont
  {Kalda}(2011)}]{Mandre11}%
  \BibitemOpen
  \bibfield  {author} {\bibinfo {author} {\bibfnamefont {I.}~\bibnamefont
  {Mandre}}\ and\ \bibinfo {author} {\bibfnamefont {J.}~\bibnamefont {Kalda}},\
  }\href {\doibase 10.1140/epjb/e2011-20386-4} {\bibfield  {journal} {\bibinfo
  {journal} {Eur. Phys. J. B}\ }\textbf {\bibinfo {volume} {83}},\ \bibinfo
  {pages} {107} (\bibinfo {year} {2011})}\BibitemShut {NoStop}%
\bibitem [{\citenamefont {Duplantier}(2000)}]{Duplantier00}%
  \BibitemOpen
  \bibfield  {author} {\bibinfo {author} {\bibfnamefont {B.}~\bibnamefont
  {Duplantier}},\ }\href {\doibase 10.1103/PhysRevLett.84.1363} {\bibfield
  {journal} {\bibinfo  {journal} {Phys. Rev. Lett.}\ }\textbf {\bibinfo
  {volume} {84}},\ \bibinfo {pages} {1363} (\bibinfo {year}
  {2000})}\BibitemShut {NoStop}%
\bibitem [{\citenamefont {Grassberger}(1983)}]{Grassberger83}%
  \BibitemOpen
  \bibfield  {author} {\bibinfo {author} {\bibfnamefont {P.}~\bibnamefont
  {Grassberger}},\ }\href {\doibase 10.1016/0025-5564(82)90036-0} {\bibfield
  {journal} {\bibinfo  {journal} {Math. Boisci.}\ }\textbf {\bibinfo {volume}
  {63}},\ \bibinfo {pages} {157} (\bibinfo {year} {1983})}\BibitemShut
  {NoStop}%
\bibitem [{\citenamefont {Mollison}(1977)}]{Mollison77}%
  \BibitemOpen
  \bibfield  {author} {\bibinfo {author} {\bibfnamefont {D.}~\bibnamefont
  {Mollison}},\ }\href@noop {} {\bibfield  {journal} {\bibinfo  {journal} {J.
  R. Statist. Soc.}\ }\textbf {\bibinfo {volume} {39}},\ \bibinfo {pages} {283}
  (\bibinfo {year} {1977})}\BibitemShut {NoStop}%
\bibitem [{\citenamefont {Murray}(2005)}]{Murray05}%
  \BibitemOpen
  \bibfield  {author} {\bibinfo {author} {\bibfnamefont {J.~D.}\ \bibnamefont
  {Murray}},\ }\href@noop {} {\emph {\bibinfo {title} {Mathematical
  Biology}}},\ \bibinfo {edition} {3rd}\ ed.\ (\bibinfo  {publisher}
  {Springer},\ \bibinfo {address} {Berlin},\ \bibinfo {year}
  {2005})\BibitemShut {NoStop}%
\bibitem [{\citenamefont {Dyson}(1969)}]{Dyson69}%
  \BibitemOpen
  \bibfield  {author} {\bibinfo {author} {\bibfnamefont {F.~J.}\ \bibnamefont
  {Dyson}},\ }\href@noop {} {\bibfield  {journal} {\bibinfo  {journal} {Commun.
  Math. Phys.}\ }\textbf {\bibinfo {volume} {91}},\ \bibinfo {pages} {212}
  (\bibinfo {year} {1969})}\BibitemShut {NoStop}%
\bibitem [{\citenamefont {Anderson}\ \emph {et~al.}(1970)\citenamefont
  {Anderson}, \citenamefont {Yuval},\ and\ \citenamefont
  {Hamann}}]{Anderson70}%
  \BibitemOpen
  \bibfield  {author} {\bibinfo {author} {\bibfnamefont {P.~W.}\ \bibnamefont
  {Anderson}}, \bibinfo {author} {\bibfnamefont {G.}~\bibnamefont {Yuval}}, \
  and\ \bibinfo {author} {\bibfnamefont {D.~R.}\ \bibnamefont {Hamann}},\
  }\href@noop {} {\bibfield  {journal} {\bibinfo  {journal} {Phys. Rev. B}\
  }\textbf {\bibinfo {volume} {1}},\ \bibinfo {pages} {4464} (\bibinfo {year}
  {1970})}\BibitemShut {NoStop}%
\bibitem [{\citenamefont {Thouless}(1969)}]{Thouless69}%
  \BibitemOpen
  \bibfield  {author} {\bibinfo {author} {\bibfnamefont {D.~J.}\ \bibnamefont
  {Thouless}},\ }\href@noop {} {\bibfield  {journal} {\bibinfo  {journal}
  {Phys. Rev.}\ }\textbf {\bibinfo {volume} {187}},\ \bibinfo {pages} {732}
  (\bibinfo {year} {1969})}\BibitemShut {NoStop}%
\bibitem [{\citenamefont {Kosterlitz}\ and\ \citenamefont
  {Thouless}(1973)}]{Kosterlitz73}%
  \BibitemOpen
  \bibfield  {author} {\bibinfo {author} {\bibfnamefont {J.~M.}\ \bibnamefont
  {Kosterlitz}}\ and\ \bibinfo {author} {\bibfnamefont {D.~J.}\ \bibnamefont
  {Thouless}},\ }\href@noop {} {\bibfield  {journal} {\bibinfo  {journal} {J.
  Phys. C}\ }\textbf {\bibinfo {volume} {6}},\ \bibinfo {pages} {1181}
  (\bibinfo {year} {1973})}\BibitemShut {NoStop}%
\bibitem [{\citenamefont {Aizenman}\ and\ \citenamefont
  {Newman}(1986)}]{Aizenman86}%
  \BibitemOpen
  \bibfield  {author} {\bibinfo {author} {\bibfnamefont {M.}~\bibnamefont
  {Aizenman}}\ and\ \bibinfo {author} {\bibfnamefont {C.~M.}\ \bibnamefont
  {Newman}},\ }\href@noop {} {\bibfield  {journal} {\bibinfo  {journal}
  {Commun. Math. Phys.}\ }\textbf {\bibinfo {volume} {107}},\ \bibinfo {pages}
  {611} (\bibinfo {year} {1986})}\BibitemShut {NoStop}%
\bibitem [{\citenamefont {Grassberger}(2013{\natexlab{a}})}]{Grassberger13}%
  \BibitemOpen
  \bibfield  {author} {\bibinfo {author} {\bibfnamefont {P.}~\bibnamefont
  {Grassberger}},\ }\href {\doibase 10.1088/1742-5468/2013/04/P04004}
  {\bibfield  {journal} {\bibinfo  {journal} {J. Stat. Mech.}\ ,\ \bibinfo
  {pages} {P04004}} (\bibinfo {year} {2013}{\natexlab{a}})}\BibitemShut
  {NoStop}%
\bibitem [{\citenamefont {Boettcher}\ \emph {et~al.}(2012)\citenamefont
  {Boettcher}, \citenamefont {Singh},\ and\ \citenamefont
  {Ziff}}]{Boettcher11}%
  \BibitemOpen
  \bibfield  {author} {\bibinfo {author} {\bibfnamefont {S.}~\bibnamefont
  {Boettcher}}, \bibinfo {author} {\bibfnamefont {V.}~\bibnamefont {Singh}}, \
  and\ \bibinfo {author} {\bibfnamefont {R.~M.}\ \bibnamefont {Ziff}},\ }\href
  {\doibase 10.1038/ncomms1774} {\bibfield  {journal} {\bibinfo  {journal}
  {Nat. Commun.}\ }\textbf {\bibinfo {volume} {3}},\ \bibinfo {pages} {787}
  (\bibinfo {year} {2012})}\BibitemShut {NoStop}%
\bibitem [{\citenamefont {Linder}\ \emph {et~al.}(2008)\citenamefont {Linder},
  \citenamefont {{Tran-Gia}}, \citenamefont {Dahmen},\ and\ \citenamefont
  {Hinrichsen}}]{Linder08}%
  \BibitemOpen
  \bibfield  {author} {\bibinfo {author} {\bibfnamefont {F.}~\bibnamefont
  {Linder}}, \bibinfo {author} {\bibfnamefont {J.}~\bibnamefont {{Tran-Gia}}},
  \bibinfo {author} {\bibfnamefont {S.~R.}\ \bibnamefont {Dahmen}}, \ and\
  \bibinfo {author} {\bibfnamefont {H.}~\bibnamefont {Hinrichsen}},\ }\href
  {\doibase 10.1088/1751-8113/41/18/185005} {\bibfield  {journal} {\bibinfo
  {journal} {J. Phys. A}\ }\textbf {\bibinfo {volume} {41}},\ \bibinfo {pages}
  {185005} (\bibinfo {year} {2008})}\BibitemShut {NoStop}%
\bibitem [{\citenamefont {Luijten}\ and\ \citenamefont
  {Bl\"ote}(2002)}]{Luijten02}%
  \BibitemOpen
  \bibfield  {author} {\bibinfo {author} {\bibfnamefont {E.}~\bibnamefont
  {Luijten}}\ and\ \bibinfo {author} {\bibfnamefont {H.~W.~J.}\ \bibnamefont
  {Bl\"ote}},\ }\href {\doibase 10.1103/PhysRevLett.89.025703} {\bibfield
  {journal} {\bibinfo  {journal} {Phys. Rev. Lett.}\ }\textbf {\bibinfo
  {volume} {89}},\ \bibinfo {pages} {025703} (\bibinfo {year}
  {2002})}\BibitemShut {NoStop}%
\bibitem [{\citenamefont {Grassberger}(2013{\natexlab{b}})}]{Grassberger13b}%
  \BibitemOpen
  \bibfield  {author} {\bibinfo {author} {\bibfnamefont {P.}~\bibnamefont
  {Grassberger}},\ }\href {\doibase 10.1007/s10955-013-0824-7} {\bibfield
  {journal} {\bibinfo  {journal} {J. Stat. Phys.}\ }\textbf {\bibinfo {volume}
  {153}},\ \bibinfo {pages} {289} (\bibinfo {year}
  {2013}{\natexlab{b}})}\BibitemShut {NoStop}%
\bibitem [{\citenamefont {Picco}()}]{Picco12}%
  \BibitemOpen
  \bibfield  {author} {\bibinfo {author} {\bibfnamefont {M.}~\bibnamefont
  {Picco}},\ }\href@noop {} {}\Eprint {http://arxiv.org/abs/arXiv:1207.1018}
  {arXiv:1207.1018} \BibitemShut {NoStop}%
\bibitem [{\citenamefont {Blanchard}\ \emph {et~al.}(2013)\citenamefont
  {Blanchard}, \citenamefont {Picco},\ and\ \citenamefont
  {Rajabpour}}]{Blanchard12}%
  \BibitemOpen
  \bibfield  {author} {\bibinfo {author} {\bibfnamefont {T.}~\bibnamefont
  {Blanchard}}, \bibinfo {author} {\bibfnamefont {M.}~\bibnamefont {Picco}}, \
  and\ \bibinfo {author} {\bibfnamefont {M.~A.}\ \bibnamefont {Rajabpour}},\
  }\href {\doibase 10.1209/0295-5075/101/56003} {\bibfield  {journal} {\bibinfo
   {journal} {EPL}\ }\textbf {\bibinfo {volume} {101}},\ \bibinfo {pages}
  {56003} (\bibinfo {year} {2013})}\BibitemShut {NoStop}%
\bibitem [{\citenamefont {Ara\'ujo}\ and\ \citenamefont
  {Herrmann}(2010)}]{Araujo10}%
  \BibitemOpen
  \bibfield  {author} {\bibinfo {author} {\bibfnamefont {N.~A.~M.}\
  \bibnamefont {Ara\'ujo}}\ and\ \bibinfo {author} {\bibfnamefont {H.~J.}\
  \bibnamefont {Herrmann}},\ }\href {\doibase 10.1103/PhysRevLett.105.035701}
  {\bibfield  {journal} {\bibinfo  {journal} {Phys. Rev. Lett.}\ }\textbf
  {\bibinfo {volume} {105}},\ \bibinfo {pages} {035701} (\bibinfo {year}
  {2010})}\BibitemShut {NoStop}%
\bibitem [{\citenamefont {Cho}\ \emph {et~al.}(2013)\citenamefont {Cho},
  \citenamefont {Hwang}, \citenamefont {Herrmann},\ and\ \citenamefont
  {Kahng}}]{Cho13}%
  \BibitemOpen
  \bibfield  {author} {\bibinfo {author} {\bibfnamefont {Y.~S.}\ \bibnamefont
  {Cho}}, \bibinfo {author} {\bibfnamefont {S.}~\bibnamefont {Hwang}}, \bibinfo
  {author} {\bibfnamefont {H.~J.}\ \bibnamefont {Herrmann}}, \ and\ \bibinfo
  {author} {\bibfnamefont {B.}~\bibnamefont {Kahng}},\ }\href {\doibase
  10.1126/science.1230813} {\bibfield  {journal} {\bibinfo  {journal}
  {Science}\ }\textbf {\bibinfo {volume} {339}},\ \bibinfo {pages} {1185}
  (\bibinfo {year} {2013})}\BibitemShut {NoStop}%
\bibitem [{\citenamefont {{da Costa}}\ \emph {et~al.}(2010)\citenamefont {{da
  Costa}}, \citenamefont {Dorogovtsev}, \citenamefont {Goltsev},\ and\
  \citenamefont {Mendes}}]{daCosta10}%
  \BibitemOpen
  \bibfield  {author} {\bibinfo {author} {\bibfnamefont {R.~A.}\ \bibnamefont
  {{da Costa}}}, \bibinfo {author} {\bibfnamefont {S.~N.}\ \bibnamefont
  {Dorogovtsev}}, \bibinfo {author} {\bibfnamefont {A.~V.}\ \bibnamefont
  {Goltsev}}, \ and\ \bibinfo {author} {\bibfnamefont {J.~F.~F.}\ \bibnamefont
  {Mendes}},\ }\href {\doibase 10.1103/PhysRevLett.105.255701} {\bibfield
  {journal} {\bibinfo  {journal} {Phys. Rev. Lett.}\ }\textbf {\bibinfo
  {volume} {105}},\ \bibinfo {pages} {255701} (\bibinfo {year}
  {2010})}\BibitemShut {NoStop}%
\bibitem [{\citenamefont {Riordan}\ and\ \citenamefont
  {Warnke}(2011)}]{Riordan11}%
  \BibitemOpen
  \bibfield  {author} {\bibinfo {author} {\bibfnamefont {O.}~\bibnamefont
  {Riordan}}\ and\ \bibinfo {author} {\bibfnamefont {L.}~\bibnamefont
  {Warnke}},\ }\href {\doibase 10.1126/science.1206241} {\bibfield  {journal}
  {\bibinfo  {journal} {Science}\ }\textbf {\bibinfo {volume} {333}},\ \bibinfo
  {pages} {322} (\bibinfo {year} {2011})}\BibitemShut {NoStop}%
\bibitem [{\citenamefont {Grassberger}\ \emph {et~al.}(2011)\citenamefont
  {Grassberger}, \citenamefont {Christensen}, \citenamefont {Bizhani},
  \citenamefont {Son},\ and\ \citenamefont {Paczuski}}]{Grassberger11}%
  \BibitemOpen
  \bibfield  {author} {\bibinfo {author} {\bibfnamefont {P.}~\bibnamefont
  {Grassberger}}, \bibinfo {author} {\bibfnamefont {C.}~\bibnamefont
  {Christensen}}, \bibinfo {author} {\bibfnamefont {G.}~\bibnamefont
  {Bizhani}}, \bibinfo {author} {\bibfnamefont {S.-W.}\ \bibnamefont {Son}}, \
  and\ \bibinfo {author} {\bibfnamefont {M.}~\bibnamefont {Paczuski}},\
  }\href@noop {} {\bibfield  {journal} {\bibinfo  {journal} {Phys. Rev. Lett.}\
  }\textbf {\bibinfo {volume} {106}},\ \bibinfo {pages} {225701} (\bibinfo
  {year} {2011})}\BibitemShut {NoStop}%
\bibitem [{\citenamefont {Bizhani}\ \emph
  {et~al.}(2011{\natexlab{a}})\citenamefont {Bizhani}, \citenamefont {Sood},
  \citenamefont {Paczuski},\ and\ \citenamefont {Grassberger}}]{Bizhani11}%
  \BibitemOpen
  \bibfield  {author} {\bibinfo {author} {\bibfnamefont {G.}~\bibnamefont
  {Bizhani}}, \bibinfo {author} {\bibfnamefont {V.}~\bibnamefont {Sood}},
  \bibinfo {author} {\bibfnamefont {M.}~\bibnamefont {Paczuski}}, \ and\
  \bibinfo {author} {\bibfnamefont {P.}~\bibnamefont {Grassberger}},\
  }\href@noop {} {\bibfield  {journal} {\bibinfo  {journal} {Phys. Rev. E}\
  }\textbf {\bibinfo {volume} {83}},\ \bibinfo {pages} {036110} (\bibinfo
  {year} {2011}{\natexlab{a}})}\BibitemShut {NoStop}%
\bibitem [{\citenamefont {Son}\ \emph {et~al.}(2011)\citenamefont {Son},
  \citenamefont {Bizhani}, \citenamefont {Christensen}, \citenamefont
  {Grassberger},\ and\ \citenamefont {Paczuski}}]{Son11}%
  \BibitemOpen
  \bibfield  {author} {\bibinfo {author} {\bibfnamefont {S.-W.}\ \bibnamefont
  {Son}}, \bibinfo {author} {\bibfnamefont {G.}~\bibnamefont {Bizhani}},
  \bibinfo {author} {\bibfnamefont {C.}~\bibnamefont {Christensen}}, \bibinfo
  {author} {\bibfnamefont {P.}~\bibnamefont {Grassberger}}, \ and\ \bibinfo
  {author} {\bibfnamefont {M.}~\bibnamefont {Paczuski}},\ }\href@noop {}
  {\bibfield  {journal} {\bibinfo  {journal} {Europhys. Lett.}\ }\textbf
  {\bibinfo {volume} {95}},\ \bibinfo {pages} {58007} (\bibinfo {year}
  {2011})}\BibitemShut {NoStop}%
\bibitem [{\citenamefont {Christensen}\ \emph {et~al.}(2012)\citenamefont
  {Christensen}, \citenamefont {Bizhani}, \citenamefont {Son}, \citenamefont
  {Paczuski},\ and\ \citenamefont {Grassberger}}]{Christensen11}%
  \BibitemOpen
  \bibfield  {author} {\bibinfo {author} {\bibfnamefont {C.}~\bibnamefont
  {Christensen}}, \bibinfo {author} {\bibfnamefont {G.}~\bibnamefont
  {Bizhani}}, \bibinfo {author} {\bibfnamefont {S.-W.}\ \bibnamefont {Son}},
  \bibinfo {author} {\bibfnamefont {M.}~\bibnamefont {Paczuski}}, \ and\
  \bibinfo {author} {\bibfnamefont {P.}~\bibnamefont {Grassberger}},\
  }\href@noop {} {\bibfield  {journal} {\bibinfo  {journal} {EPL}\ }\textbf
  {\bibinfo {volume} {97}},\ \bibinfo {pages} {16004} (\bibinfo {year}
  {2012})}\BibitemShut {NoStop}%
\bibitem [{\citenamefont {Bizhani}\ \emph
  {et~al.}(2011{\natexlab{b}})\citenamefont {Bizhani}, \citenamefont
  {Grassberger},\ and\ \citenamefont {Paczuski}}]{Bizhani11b}%
  \BibitemOpen
  \bibfield  {author} {\bibinfo {author} {\bibfnamefont {G.}~\bibnamefont
  {Bizhani}}, \bibinfo {author} {\bibfnamefont {P.}~\bibnamefont
  {Grassberger}}, \ and\ \bibinfo {author} {\bibfnamefont {M.}~\bibnamefont
  {Paczuski}},\ }\href@noop {} {\bibfield  {journal} {\bibinfo  {journal}
  {Phys. Rev. E}\ }\textbf {\bibinfo {volume} {84}},\ \bibinfo {pages} {066111}
  (\bibinfo {year} {2011}{\natexlab{b}})}\BibitemShut {NoStop}%
\bibitem [{\citenamefont {Lau}\ \emph {et~al.}(2012)\citenamefont {Lau},
  \citenamefont {Paczuski},\ and\ \citenamefont {Grassberger}}]{Lau12}%
  \BibitemOpen
  \bibfield  {author} {\bibinfo {author} {\bibfnamefont {H.~W.}\ \bibnamefont
  {Lau}}, \bibinfo {author} {\bibfnamefont {M.}~\bibnamefont {Paczuski}}, \
  and\ \bibinfo {author} {\bibfnamefont {P.}~\bibnamefont {Grassberger}},\
  }\href {\doibase 10.1103/PhysRevE.86.011118} {\bibfield  {journal} {\bibinfo
  {journal} {Phys. Rev. E}\ }\textbf {\bibinfo {volume} {86}},\ \bibinfo
  {pages} {011118} (\bibinfo {year} {2012})}\BibitemShut {NoStop}%
\bibitem [{\citenamefont {Callaway}\ \emph {et~al.}(2001)\citenamefont
  {Callaway}, \citenamefont {Hopcroft}, \citenamefont {Kleinberg},
  \citenamefont {Newman},\ and\ \citenamefont {Strogatz}}]{Callaway01}%
  \BibitemOpen
  \bibfield  {author} {\bibinfo {author} {\bibfnamefont {D.~S.}\ \bibnamefont
  {Callaway}}, \bibinfo {author} {\bibfnamefont {J.~E.}\ \bibnamefont
  {Hopcroft}}, \bibinfo {author} {\bibfnamefont {J.~M.}\ \bibnamefont
  {Kleinberg}}, \bibinfo {author} {\bibfnamefont {M.~E.~J.}\ \bibnamefont
  {Newman}}, \ and\ \bibinfo {author} {\bibfnamefont {S.~H.}\ \bibnamefont
  {Strogatz}},\ }\href {\doibase 10.1103/PhysRevE.64.041902} {\bibfield
  {journal} {\bibinfo  {journal} {Phys. Rev. E}\ }\textbf {\bibinfo {volume}
  {64}},\ \bibinfo {pages} {041902} (\bibinfo {year} {2001})}\BibitemShut
  {NoStop}%
\bibitem [{\citenamefont {Barab\'asi}\ and\ \citenamefont
  {Albert}(1999)}]{Barabasi99}%
  \BibitemOpen
  \bibfield  {author} {\bibinfo {author} {\bibfnamefont {A.-L.}\ \bibnamefont
  {Barab\'asi}}\ and\ \bibinfo {author} {\bibfnamefont {R.}~\bibnamefont
  {Albert}},\ }\href@noop {} {\bibfield  {journal} {\bibinfo  {journal}
  {Science}\ }\textbf {\bibinfo {volume} {286}},\ \bibinfo {pages} {509}
  (\bibinfo {year} {1999})}\BibitemShut {NoStop}%
\bibitem [{\citenamefont {Chalupa}\ \emph {et~al.}(1979)\citenamefont
  {Chalupa}, \citenamefont {Leath},\ and\ \citenamefont {Reich}}]{Chalupa79}%
  \BibitemOpen
  \bibfield  {author} {\bibinfo {author} {\bibfnamefont {J.}~\bibnamefont
  {Chalupa}}, \bibinfo {author} {\bibfnamefont {P.~L.}\ \bibnamefont {Leath}},
  \ and\ \bibinfo {author} {\bibfnamefont {G.~R.}\ \bibnamefont {Reich}},\
  }\href@noop {} {\bibfield  {journal} {\bibinfo  {journal} {J. Phys. C}\
  }\textbf {\bibinfo {volume} {12}},\ \bibinfo {pages} {L31} (\bibinfo {year}
  {1979})}\BibitemShut {NoStop}%
\bibitem [{\citenamefont {Adler}(1991)}]{Adler91}%
  \BibitemOpen
  \bibfield  {author} {\bibinfo {author} {\bibfnamefont {J.}~\bibnamefont
  {Adler}},\ }\href@noop {} {\bibfield  {journal} {\bibinfo  {journal} {Physica
  A}\ }\textbf {\bibinfo {volume} {171}},\ \bibinfo {pages} {453} (\bibinfo
  {year} {1991})}\BibitemShut {NoStop}%
\bibitem [{\citenamefont {Dodds}\ and\ \citenamefont {Watts}(2004)}]{Dodds04}%
  \BibitemOpen
  \bibfield  {author} {\bibinfo {author} {\bibfnamefont {P.~S.}\ \bibnamefont
  {Dodds}}\ and\ \bibinfo {author} {\bibfnamefont {D.~J.}\ \bibnamefont
  {Watts}},\ }\href@noop {} {\bibfield  {journal} {\bibinfo  {journal} {Phys.
  Rev. Lett.}\ }\textbf {\bibinfo {volume} {92}},\ \bibinfo {pages} {218701}
  (\bibinfo {year} {2004})}\BibitemShut {NoStop}%
\bibitem [{\citenamefont {Dodds}\ and\ \citenamefont {Watts}(2005)}]{Dodds05}%
  \BibitemOpen
  \bibfield  {author} {\bibinfo {author} {\bibfnamefont {P.~S.}\ \bibnamefont
  {Dodds}}\ and\ \bibinfo {author} {\bibfnamefont {D.~J.}\ \bibnamefont
  {Watts}},\ }\href@noop {} {\bibfield  {journal} {\bibinfo  {journal} {J.
  Theor. Biology}\ }\textbf {\bibinfo {volume} {42}},\ \bibinfo {pages} {232}
  (\bibinfo {year} {2005})}\BibitemShut {NoStop}%
\bibitem [{\citenamefont {Janssen}\ \emph {et~al.}(2004)\citenamefont
  {Janssen}, \citenamefont {M\"uller},\ and\ \citenamefont
  {Stenull}}]{Janssen04}%
  \BibitemOpen
  \bibfield  {author} {\bibinfo {author} {\bibfnamefont {H.-K.}\ \bibnamefont
  {Janssen}}, \bibinfo {author} {\bibfnamefont {M.}~\bibnamefont {M\"uller}}, \
  and\ \bibinfo {author} {\bibfnamefont {O.}~\bibnamefont {Stenull}},\
  }\href@noop {} {\bibfield  {journal} {\bibinfo  {journal} {Phys. Rev. E}\
  }\textbf {\bibinfo {volume} {70}},\ \bibinfo {pages} {026114} (\bibinfo
  {year} {2004})}\BibitemShut {NoStop}%
\bibitem [{\citenamefont {Bizhani}\ \emph {et~al.}(2012)\citenamefont
  {Bizhani}, \citenamefont {Paczuski},\ and\ \citenamefont
  {Grassberger}}]{Bizhani12}%
  \BibitemOpen
  \bibfield  {author} {\bibinfo {author} {\bibfnamefont {G.}~\bibnamefont
  {Bizhani}}, \bibinfo {author} {\bibfnamefont {M.}~\bibnamefont {Paczuski}}, \
  and\ \bibinfo {author} {\bibfnamefont {P.}~\bibnamefont {Grassberger}},\
  }\href {\doibase 10.1103/PhysRevE.86.011128} {\bibfield  {journal} {\bibinfo
  {journal} {Phys. Rev. E}\ }\textbf {\bibinfo {volume} {86}},\ \bibinfo
  {pages} {011128} (\bibinfo {year} {2012})}\BibitemShut {NoStop}%
\bibitem [{\citenamefont {Barab\'asi}\ and\ \citenamefont
  {Stanley}(1995)}]{Barabasi95}%
  \BibitemOpen
  \bibfield  {author} {\bibinfo {author} {\bibfnamefont {A.-L.}\ \bibnamefont
  {Barab\'asi}}\ and\ \bibinfo {author} {\bibfnamefont {H.~E.}\ \bibnamefont
  {Stanley}},\ }\href@noop {} {\emph {\bibinfo {title} {Fractal Concepts in
  Surface Growth}}}\ (\bibinfo  {publisher} {Cambridge University Press},\
  \bibinfo {address} {New York},\ \bibinfo {year} {1995})\BibitemShut {NoStop}%
\bibitem [{\citenamefont {Goltsev}\ \emph {et~al.}(2006)\citenamefont
  {Goltsev}, \citenamefont {Dorogovtsev},\ and\ \citenamefont
  {Mendes}}]{Goltsev06}%
  \BibitemOpen
  \bibfield  {author} {\bibinfo {author} {\bibfnamefont {A.~V.}\ \bibnamefont
  {Goltsev}}, \bibinfo {author} {\bibfnamefont {S.~N.}\ \bibnamefont
  {Dorogovtsev}}, \ and\ \bibinfo {author} {\bibfnamefont {J.~F.~F.}\
  \bibnamefont {Mendes}},\ }\href@noop {} {\bibfield  {journal} {\bibinfo
  {journal} {Phys. Rev. E}\ }\textbf {\bibinfo {volume} {73}},\ \bibinfo
  {pages} {056101} (\bibinfo {year} {2006})}\BibitemShut {NoStop}%
\bibitem [{\citenamefont {{Le Doussal}}\ \emph {et~al.}(2002)\citenamefont {{Le
  Doussal}}, \citenamefont {Wiese},\ and\ \citenamefont {Chauve}}]{Doussal02}%
  \BibitemOpen
  \bibfield  {author} {\bibinfo {author} {\bibfnamefont {P.}~\bibnamefont {{Le
  Doussal}}}, \bibinfo {author} {\bibfnamefont {K.~J.}\ \bibnamefont {Wiese}},
  \ and\ \bibinfo {author} {\bibfnamefont {P.}~\bibnamefont {Chauve}},\
  }\href@noop {} {\bibfield  {journal} {\bibinfo  {journal} {Phys. Rev. B}\
  }\textbf {\bibinfo {volume} {66}},\ \bibinfo {pages} {174201} (\bibinfo
  {year} {2002})}\BibitemShut {NoStop}%
\bibitem [{\citenamefont {Drossel}\ and\ \citenamefont
  {Dahmen}(1998)}]{Drossel98}%
  \BibitemOpen
  \bibfield  {author} {\bibinfo {author} {\bibfnamefont {B.}~\bibnamefont
  {Drossel}}\ and\ \bibinfo {author} {\bibfnamefont {K.}~\bibnamefont
  {Dahmen}},\ }\href@noop {} {\bibfield  {journal} {\bibinfo  {journal} {Eur.
  Phys. J. B}\ }\textbf {\bibinfo {volume} {3}},\ \bibinfo {pages} {485}
  (\bibinfo {year} {1998})}\BibitemShut {NoStop}%
\bibitem [{\citenamefont {Zhou}\ and\ \citenamefont {Zheng}(2010)}]{Zhou10}%
  \BibitemOpen
  \bibfield  {author} {\bibinfo {author} {\bibfnamefont {N.~J.}\ \bibnamefont
  {Zhou}}\ and\ \bibinfo {author} {\bibfnamefont {B.}~\bibnamefont {Zheng}},\
  }\href@noop {} {\bibfield  {journal} {\bibinfo  {journal} {Phys. Rev. E}\
  }\textbf {\bibinfo {volume} {82}},\ \bibinfo {pages} {031139} (\bibinfo
  {year} {2010})}\BibitemShut {NoStop}%
\bibitem [{\citenamefont {Qin}\ \emph {et~al.}(2012)\citenamefont {Qin},
  \citenamefont {Zheng},\ and\ \citenamefont {Zhou}}]{Qin12}%
  \BibitemOpen
  \bibfield  {author} {\bibinfo {author} {\bibfnamefont {X.~P.}\ \bibnamefont
  {Qin}}, \bibinfo {author} {\bibfnamefont {B.}~\bibnamefont {Zheng}}, \ and\
  \bibinfo {author} {\bibfnamefont {N.~J.}\ \bibnamefont {Zhou}},\ }\href@noop
  {} {\bibfield  {journal} {\bibinfo  {journal} {J. Phys. A}\ }\textbf
  {\bibinfo {volume} {45}},\ \bibinfo {pages} {345005} (\bibinfo {year}
  {2012})}\BibitemShut {NoStop}%
\bibitem [{\citenamefont {Aizenman}\ and\ \citenamefont
  {Wehr}(1989)}]{Aizenman89}%
  \BibitemOpen
  \bibfield  {author} {\bibinfo {author} {\bibfnamefont {M.}~\bibnamefont
  {Aizenman}}\ and\ \bibinfo {author} {\bibfnamefont {J.}~\bibnamefont
  {Wehr}},\ }\href@noop {} {\bibfield  {journal} {\bibinfo  {journal} {Phys.
  Rev. Lett.}\ }\textbf {\bibinfo {volume} {62}},\ \bibinfo {pages} {2503}
  (\bibinfo {year} {1989})}\BibitemShut {NoStop}%
\bibitem [{\citenamefont {Sethna}\ \emph {et~al.}()\citenamefont {Sethna},
  \citenamefont {Dahmen},\ and\ \citenamefont {Perkovi\'c}}]{Sethna06}%
  \BibitemOpen
  \bibfield  {author} {\bibinfo {author} {\bibfnamefont {J.~P.}\ \bibnamefont
  {Sethna}}, \bibinfo {author} {\bibfnamefont {K.~A.}\ \bibnamefont {Dahmen}},
  \ and\ \bibinfo {author} {\bibfnamefont {O.}~\bibnamefont {Perkovi\'c}},\
  }\href@noop {} {}\Eprint {http://arxiv.org/abs/arXiv:cond-mat/0406320}
  {arXiv:cond-mat/0406320} \BibitemShut {NoStop}%
\bibitem [{\citenamefont {Singer}(2009)}]{Singer09}%
  \BibitemOpen
  \bibfield  {author} {\bibinfo {author} {\bibfnamefont {M.}~\bibnamefont
  {Singer}},\ }\href@noop {} {\emph {\bibinfo {title} {Introduction to
  Syndemics: A Critical Systems Approach to Public and Community Health}}}\
  (\bibinfo  {publisher} {John Wiley \& Sons},\ \bibinfo {address} {Hoboken},\
  \bibinfo {year} {2009})\BibitemShut {NoStop}%
\bibitem [{\citenamefont {Brundage}\ and\ \citenamefont
  {Shanks}(2008)}]{Brundage08}%
  \BibitemOpen
  \bibfield  {author} {\bibinfo {author} {\bibfnamefont {J.~F.}\ \bibnamefont
  {Brundage}}\ and\ \bibinfo {author} {\bibfnamefont {J.~D.}\ \bibnamefont
  {Shanks}},\ }\href@noop {} {\bibfield  {journal} {\bibinfo  {journal}
  {Emerging Infectious Diseases}\ }\textbf {\bibinfo {volume} {14}},\ \bibinfo
  {pages} {1193} (\bibinfo {year} {2008})}\BibitemShut {NoStop}%
\bibitem [{\citenamefont {Oei}\ and\ \citenamefont {Nishiura}(2012)}]{Oei12}%
  \BibitemOpen
  \bibfield  {author} {\bibinfo {author} {\bibfnamefont {W.}~\bibnamefont
  {Oei}}\ and\ \bibinfo {author} {\bibfnamefont {H.}~\bibnamefont {Nishiura}},\
  }\href@noop {} {\bibfield  {journal} {\bibinfo  {journal} {Comput. Math.
  Methods in Med.}\ }\textbf {\bibinfo {volume} {2012}},\ \bibinfo {pages}
  {124861} (\bibinfo {year} {2012})}\BibitemShut {NoStop}%
\bibitem [{\citenamefont {Sulkowski}(2008)}]{Sulkowski08}%
  \BibitemOpen
  \bibfield  {author} {\bibinfo {author} {\bibfnamefont {M.~S.}\ \bibnamefont
  {Sulkowski}},\ }\href@noop {} {\bibfield  {journal} {\bibinfo  {journal} {J.
  Hepatology}\ }\textbf {\bibinfo {volume} {48}},\ \bibinfo {pages} {353}
  (\bibinfo {year} {2008})}\BibitemShut {NoStop}%
\bibitem [{\citenamefont {Sharma}\ \emph {et~al.}(2005)\citenamefont {Sharma},
  \citenamefont {Mohan},\ and\ \citenamefont {Kadhivaran}}]{Sharma05}%
  \BibitemOpen
  \bibfield  {author} {\bibinfo {author} {\bibfnamefont {S.~K.}\ \bibnamefont
  {Sharma}}, \bibinfo {author} {\bibfnamefont {A.}~\bibnamefont {Mohan}}, \
  and\ \bibinfo {author} {\bibfnamefont {S.}~\bibnamefont {Kadhivaran}},\
  }\href@noop {} {\bibfield  {journal} {\bibinfo  {journal} {Indian J. Med.
  Res.}\ }\textbf {\bibinfo {volume} {121}},\ \bibinfo {pages} {550} (\bibinfo
  {year} {2005})}\BibitemShut {NoStop}%
\bibitem [{\citenamefont {Chen}\ \emph
  {et~al.}(2013{\natexlab{a}})\citenamefont {Chen}, \citenamefont {Cai},
  \citenamefont {Ghanbarnejad},\ and\ \citenamefont {Grassberger}}]{Chen13c}%
  \BibitemOpen
  \bibfield  {author} {\bibinfo {author} {\bibfnamefont {L.}~\bibnamefont
  {Chen}}, \bibinfo {author} {\bibfnamefont {W.}~\bibnamefont {Cai}}, \bibinfo
  {author} {\bibfnamefont {F.}~\bibnamefont {Ghanbarnejad}}, \ and\ \bibinfo
  {author} {\bibfnamefont {P.}~\bibnamefont {Grassberger}},\ }\href@noop {}
  {\bibfield  {journal} {\bibinfo  {journal} {Europhys. Lett.}\ }\textbf
  {\bibinfo {volume} {104}},\ \bibinfo {pages} {50001} (\bibinfo {year}
  {2013}{\natexlab{a}})}\BibitemShut {NoStop}%
\bibitem [{\citenamefont {Cai}\ \emph {et~al.}(2014)\citenamefont {Cai},
  \citenamefont {Chen}, \citenamefont {Ghanbarnejad},\ and\ \citenamefont
  {Grassberger}}]{Cai14}%
  \BibitemOpen
  \bibfield  {author} {\bibinfo {author} {\bibfnamefont {W.}~\bibnamefont
  {Cai}}, \bibinfo {author} {\bibfnamefont {L.}~\bibnamefont {Chen}}, \bibinfo
  {author} {\bibfnamefont {F.}~\bibnamefont {Ghanbarnejad}}, \ and\ \bibinfo
  {author} {\bibfnamefont {P.}~\bibnamefont {Grassberger}},\ }\href@noop {} {}
  (\bibinfo {year} {2014}),\ \bibinfo {note} {to be published}\BibitemShut
  {NoStop}%
\bibitem [{\citenamefont {Buldyrev}\ \emph {et~al.}(2010)\citenamefont
  {Buldyrev}, \citenamefont {Parshani}, \citenamefont {Paul}, \citenamefont
  {Stanley},\ and\ \citenamefont {Havlin}}]{Buldyrev10}%
  \BibitemOpen
  \bibfield  {author} {\bibinfo {author} {\bibfnamefont {S.~V.}\ \bibnamefont
  {Buldyrev}}, \bibinfo {author} {\bibfnamefont {R.}~\bibnamefont {Parshani}},
  \bibinfo {author} {\bibfnamefont {G.}~\bibnamefont {Paul}}, \bibinfo {author}
  {\bibfnamefont {H.~E.}\ \bibnamefont {Stanley}}, \ and\ \bibinfo {author}
  {\bibfnamefont {S.}~\bibnamefont {Havlin}},\ }\href@noop {} {\bibfield
  {journal} {\bibinfo  {journal} {Nature}\ }\textbf {\bibinfo {volume} {464}},\
  \bibinfo {pages} {1025} (\bibinfo {year} {2010})}\BibitemShut {NoStop}%
\bibitem [{\citenamefont {Parshani}\ \emph {et~al.}(2010)\citenamefont
  {Parshani}, \citenamefont {Buldyrev},\ and\ \citenamefont
  {Havlin}}]{Parshani10}%
  \BibitemOpen
  \bibfield  {author} {\bibinfo {author} {\bibfnamefont {R.}~\bibnamefont
  {Parshani}}, \bibinfo {author} {\bibfnamefont {S.~V.}\ \bibnamefont
  {Buldyrev}}, \ and\ \bibinfo {author} {\bibfnamefont {S.}~\bibnamefont
  {Havlin}},\ }\href@noop {} {\bibfield  {journal} {\bibinfo  {journal} {Phys.
  Rev. Lett.}\ }\textbf {\bibinfo {volume} {105}},\ \bibinfo {pages} {048701}
  (\bibinfo {year} {2010})}\BibitemShut {NoStop}%
\bibitem [{\citenamefont {Gao}\ \emph {et~al.}(2011)\citenamefont {Gao},
  \citenamefont {Buldyrev}, \citenamefont {Stanley},\ and\ \citenamefont
  {Havlin}}]{Gao11}%
  \BibitemOpen
  \bibfield  {author} {\bibinfo {author} {\bibfnamefont {J.}~\bibnamefont
  {Gao}}, \bibinfo {author} {\bibfnamefont {S.~V.}\ \bibnamefont {Buldyrev}},
  \bibinfo {author} {\bibfnamefont {H.~E.}\ \bibnamefont {Stanley}}, \ and\
  \bibinfo {author} {\bibfnamefont {S.}~\bibnamefont {Havlin}},\ }\href@noop {}
  {\bibfield  {journal} {\bibinfo  {journal} {Nat. Phys.}\ }\textbf {\bibinfo
  {volume} {8}},\ \bibinfo {pages} {40} (\bibinfo {year} {2011})}\BibitemShut
  {NoStop}%
\bibitem [{\citenamefont {Son}\ \emph {et~al.}(2012)\citenamefont {Son},
  \citenamefont {Bizhani}, \citenamefont {Christensen}, \citenamefont
  {Grassberger},\ and\ \citenamefont {Paczuski}}]{Son12}%
  \BibitemOpen
  \bibfield  {author} {\bibinfo {author} {\bibfnamefont {S.-W.}\ \bibnamefont
  {Son}}, \bibinfo {author} {\bibfnamefont {G.}~\bibnamefont {Bizhani}},
  \bibinfo {author} {\bibfnamefont {C.}~\bibnamefont {Christensen}}, \bibinfo
  {author} {\bibfnamefont {P.}~\bibnamefont {Grassberger}}, \ and\ \bibinfo
  {author} {\bibfnamefont {M.}~\bibnamefont {Paczuski}},\ }\href@noop {}
  {\bibfield  {journal} {\bibinfo  {journal} {Europhys. Lett.}\ }\textbf
  {\bibinfo {volume} {97}},\ \bibinfo {pages} {16006} (\bibinfo {year}
  {2012})}\BibitemShut {NoStop}%
\bibitem [{\citenamefont {Bollob\'as}(1984)}]{Bollobas84}%
  \BibitemOpen
  \bibfield  {author} {\bibinfo {author} {\bibfnamefont {B.}~\bibnamefont
  {Bollob\'as}},\ }\href@noop {} {\emph {\bibinfo {title} {Graph Theory and
  Combinatorics: Proc. Cambridge Combinatorial Conference in Honour of Paul
  Erd\H{o}s}}}\ (\bibinfo  {publisher} {Academic Press},\ \bibinfo {address}
  {New York},\ \bibinfo {year} {1984})\ p.~\bibinfo {pages} {35}\BibitemShut
  {NoStop}%
\bibitem [{\citenamefont {Dorogovtsev}\ \emph {et~al.}(2006)\citenamefont
  {Dorogovtsev}, \citenamefont {Goltsev},\ and\ \citenamefont
  {Mendes}}]{Dorogovtsev06}%
  \BibitemOpen
  \bibfield  {author} {\bibinfo {author} {\bibfnamefont {S.~N.}\ \bibnamefont
  {Dorogovtsev}}, \bibinfo {author} {\bibfnamefont {A.~V.}\ \bibnamefont
  {Goltsev}}, \ and\ \bibinfo {author} {\bibfnamefont {J.~F.~F.}\ \bibnamefont
  {Mendes}},\ }\href@noop {} {\bibfield  {journal} {\bibinfo  {journal} {Phys.
  Rev. Lett.}\ }\textbf {\bibinfo {volume} {96}},\ \bibinfo {pages} {040601}
  (\bibinfo {year} {2006})}\BibitemShut {NoStop}%
\bibitem [{\citenamefont {Ziff}(2009)}]{Ziff09}%
  \BibitemOpen
  \bibfield  {author} {\bibinfo {author} {\bibfnamefont {R.~M.}\ \bibnamefont
  {Ziff}},\ }\href {\doibase 10.1103/PhysRevLett.103.045701} {\bibfield
  {journal} {\bibinfo  {journal} {Phys. Rev. Lett.}\ }\textbf {\bibinfo
  {volume} {103}},\ \bibinfo {pages} {045701} (\bibinfo {year}
  {2009})}\BibitemShut {NoStop}%
\bibitem [{\citenamefont {Cho}\ \emph {et~al.}(2009)\citenamefont {Cho},
  \citenamefont {Kim}, \citenamefont {Park}, \citenamefont {Kahng},\ and\
  \citenamefont {Kim}}]{Cho09}%
  \BibitemOpen
  \bibfield  {author} {\bibinfo {author} {\bibfnamefont {Y.~S.}\ \bibnamefont
  {Cho}}, \bibinfo {author} {\bibfnamefont {J.~S.}\ \bibnamefont {Kim}},
  \bibinfo {author} {\bibfnamefont {J.}~\bibnamefont {Park}}, \bibinfo {author}
  {\bibfnamefont {B.}~\bibnamefont {Kahng}}, \ and\ \bibinfo {author}
  {\bibfnamefont {D.}~\bibnamefont {Kim}},\ }\href@noop {} {\bibfield
  {journal} {\bibinfo  {journal} {Phys. Rev. Lett.}\ }\textbf {\bibinfo
  {volume} {103}},\ \bibinfo {pages} {135702} (\bibinfo {year}
  {2009})}\BibitemShut {NoStop}%
\bibitem [{\citenamefont {Friedman}\ and\ \citenamefont
  {Landsberg}(2009)}]{Friedman09}%
  \BibitemOpen
  \bibfield  {author} {\bibinfo {author} {\bibfnamefont {E.~J.}\ \bibnamefont
  {Friedman}}\ and\ \bibinfo {author} {\bibfnamefont {A.~S.}\ \bibnamefont
  {Landsberg}},\ }\href@noop {} {\bibfield  {journal} {\bibinfo  {journal}
  {Phys. Rev. Lett.}\ }\textbf {\bibinfo {volume} {103}},\ \bibinfo {pages}
  {255701} (\bibinfo {year} {2009})}\BibitemShut {NoStop}%
\bibitem [{\citenamefont {D'Souza}\ and\ \citenamefont
  {Mitzenmacher}(2010)}]{DSouza10}%
  \BibitemOpen
  \bibfield  {author} {\bibinfo {author} {\bibfnamefont {R.~M.}\ \bibnamefont
  {D'Souza}}\ and\ \bibinfo {author} {\bibfnamefont {M.}~\bibnamefont
  {Mitzenmacher}},\ }\href@noop {} {\bibfield  {journal} {\bibinfo  {journal}
  {Phys. Rev. Lett.}\ }\textbf {\bibinfo {volume} {104}},\ \bibinfo {pages}
  {195702} (\bibinfo {year} {2010})}\BibitemShut {NoStop}%
\bibitem [{\citenamefont {Nagler}\ \emph {et~al.}(2011)\citenamefont {Nagler},
  \citenamefont {Levina},\ and\ \citenamefont {Timme}}]{Nagler11}%
  \BibitemOpen
  \bibfield  {author} {\bibinfo {author} {\bibfnamefont {J.}~\bibnamefont
  {Nagler}}, \bibinfo {author} {\bibfnamefont {A.}~\bibnamefont {Levina}}, \
  and\ \bibinfo {author} {\bibfnamefont {M.}~\bibnamefont {Timme}},\ }\href
  {\doibase 10.1038/NPHYS1860} {\bibfield  {journal} {\bibinfo  {journal} {Nat.
  Phys.}\ }\textbf {\bibinfo {volume} {7}},\ \bibinfo {pages} {265} (\bibinfo
  {year} {2011})}\BibitemShut {NoStop}%
\bibitem [{\citenamefont {Ziff}(2013)}]{Ziff13}%
  \BibitemOpen
  \bibfield  {author} {\bibinfo {author} {\bibfnamefont {R.~M.}\ \bibnamefont
  {Ziff}},\ }\href {\doibase 10.1126/science.1235032} {\bibfield  {journal}
  {\bibinfo  {journal} {Science}\ }\textbf {\bibinfo {volume} {339}},\ \bibinfo
  {pages} {1159} (\bibinfo {year} {2013})}\BibitemShut {NoStop}%
\bibitem [{\citenamefont {Cho}\ and\ \citenamefont {Kahng}()}]{Cho14}%
  \BibitemOpen
  \bibfield  {author} {\bibinfo {author} {\bibfnamefont {Y.~S.}\ \bibnamefont
  {Cho}}\ and\ \bibinfo {author} {\bibfnamefont {B.}~\bibnamefont {Kahng}},\
  }\href@noop {} {}\Eprint {http://arxiv.org/abs/arXiv:1404.4470}
  {arXiv:1404.4470} \BibitemShut {NoStop}%
\bibitem [{\citenamefont {Ziff}(2010)}]{Ziff10}%
  \BibitemOpen
  \bibfield  {author} {\bibinfo {author} {\bibfnamefont {R.~M.}\ \bibnamefont
  {Ziff}},\ }\href {\doibase 10.1103/PhysRevE.82.051105} {\bibfield  {journal}
  {\bibinfo  {journal} {Phys. Rev. E}\ }\textbf {\bibinfo {volume} {82}},\
  \bibinfo {pages} {051105} (\bibinfo {year} {2010})}\BibitemShut {NoStop}%
\bibitem [{\citenamefont {Cho}\ \emph {et~al.}(2010)\citenamefont {Cho},
  \citenamefont {Kahng},\ and\ \citenamefont {Kim}}]{Cho10}%
  \BibitemOpen
  \bibfield  {author} {\bibinfo {author} {\bibfnamefont {Y.~S.}\ \bibnamefont
  {Cho}}, \bibinfo {author} {\bibfnamefont {B.}~\bibnamefont {Kahng}}, \ and\
  \bibinfo {author} {\bibfnamefont {D.}~\bibnamefont {Kim}},\ }\href@noop {}
  {\bibfield  {journal} {\bibinfo  {journal} {Phys. Rev. E}\ }\textbf {\bibinfo
  {volume} {81}},\ \bibinfo {pages} {030103(R)} (\bibinfo {year}
  {2010})}\BibitemShut {NoStop}%
\bibitem [{\citenamefont {Bohman}\ \emph {et~al.}(2004)\citenamefont {Bohman},
  \citenamefont {Frieze},\ and\ \citenamefont {Wormald}}]{Bohman04}%
  \BibitemOpen
  \bibfield  {author} {\bibinfo {author} {\bibfnamefont {T.}~\bibnamefont
  {Bohman}}, \bibinfo {author} {\bibfnamefont {A.}~\bibnamefont {Frieze}}, \
  and\ \bibinfo {author} {\bibfnamefont {N.~C.}\ \bibnamefont {Wormald}},\
  }\href@noop {} {\bibfield  {journal} {\bibinfo  {journal} {Random Struct.
  Algorithms}\ }\textbf {\bibinfo {volume} {25}},\ \bibinfo {pages} {432}
  (\bibinfo {year} {2004})}\BibitemShut {NoStop}%
\bibitem [{\citenamefont {Chen}\ and\ \citenamefont {D'Souza}(2011)}]{Chen11}%
  \BibitemOpen
  \bibfield  {author} {\bibinfo {author} {\bibfnamefont {W.}~\bibnamefont
  {Chen}}\ and\ \bibinfo {author} {\bibfnamefont {R.~M.}\ \bibnamefont
  {D'Souza}},\ }\href {\doibase 10.1103/PhysRevLett.106.115701} {\bibfield
  {journal} {\bibinfo  {journal} {Phys. Rev. Lett.}\ }\textbf {\bibinfo
  {volume} {106}},\ \bibinfo {pages} {115701} (\bibinfo {year}
  {2011})}\BibitemShut {NoStop}%
\bibitem [{\citenamefont {Chen}\ \emph {et~al.}(2012)\citenamefont {Chen},
  \citenamefont {Zheng},\ and\ \citenamefont {D'Souza}}]{Chen11b}%
  \BibitemOpen
  \bibfield  {author} {\bibinfo {author} {\bibfnamefont {W.}~\bibnamefont
  {Chen}}, \bibinfo {author} {\bibfnamefont {Z.}~\bibnamefont {Zheng}}, \ and\
  \bibinfo {author} {\bibfnamefont {R.~M.}\ \bibnamefont {D'Souza}},\ }\href
  {\doibase 10.1209/0295-5075/100/66006} {\bibfield  {journal} {\bibinfo
  {journal} {EPL}\ }\textbf {\bibinfo {volume} {100}},\ \bibinfo {pages}
  {66006} (\bibinfo {year} {2012})}\BibitemShut {NoStop}%
\bibitem [{\citenamefont {Schrenk}\ \emph
  {et~al.}(2012{\natexlab{b}})\citenamefont {Schrenk}, \citenamefont {Felder},
  \citenamefont {Deflorin}, \citenamefont {Ara\'ujo}, \citenamefont {D'Souza},\
  and\ \citenamefont {Herrmann}}]{Schrenk11b}%
  \BibitemOpen
  \bibfield  {author} {\bibinfo {author} {\bibfnamefont {K.~J.}\ \bibnamefont
  {Schrenk}}, \bibinfo {author} {\bibfnamefont {A.}~\bibnamefont {Felder}},
  \bibinfo {author} {\bibfnamefont {S.}~\bibnamefont {Deflorin}}, \bibinfo
  {author} {\bibfnamefont {N.~A.~M.}\ \bibnamefont {Ara\'ujo}}, \bibinfo
  {author} {\bibfnamefont {R.~M.}\ \bibnamefont {D'Souza}}, \ and\ \bibinfo
  {author} {\bibfnamefont {H.~J.}\ \bibnamefont {Herrmann}},\ }\href {\doibase
  10.1103/PhysRevE.85.031103} {\bibfield  {journal} {\bibinfo  {journal} {Phys.
  Rev. E}\ }\textbf {\bibinfo {volume} {85}},\ \bibinfo {pages} {031103}
  (\bibinfo {year} {2012}{\natexlab{b}})}\BibitemShut {NoStop}%
\bibitem [{\citenamefont {Panagiotou}\ \emph {et~al.}(2011)\citenamefont
  {Panagiotou}, \citenamefont {Sp{\"o}hel}, \citenamefont {Steger},\ and\
  \citenamefont {Thomas}}]{Panagiotou11}%
  \BibitemOpen
  \bibfield  {author} {\bibinfo {author} {\bibfnamefont {K.}~\bibnamefont
  {Panagiotou}}, \bibinfo {author} {\bibfnamefont {R.}~\bibnamefont
  {Sp{\"o}hel}}, \bibinfo {author} {\bibfnamefont {A.}~\bibnamefont {Steger}},
  \ and\ \bibinfo {author} {\bibfnamefont {H.}~\bibnamefont {Thomas}},\ }\href
  {\doibase 10.1016/j.endm.2011.10.017} {\bibfield  {journal} {\bibinfo
  {journal} {Electron. Notes Discrete Math.}\ }\textbf {\bibinfo {volume}
  {38}},\ \bibinfo {pages} {699} (\bibinfo {year} {2011})}\BibitemShut
  {NoStop}%
\bibitem [{\citenamefont {Zhang}\ \emph {et~al.}(2013)\citenamefont {Zhang},
  \citenamefont {Wei}, \citenamefont {Guo}, \citenamefont {Zhang},\ and\
  \citenamefont {Zheng}}]{Zhang12}%
  \BibitemOpen
  \bibfield  {author} {\bibinfo {author} {\bibfnamefont {R.}~\bibnamefont
  {Zhang}}, \bibinfo {author} {\bibfnamefont {W.}~\bibnamefont {Wei}}, \bibinfo
  {author} {\bibfnamefont {B.}~\bibnamefont {Guo}}, \bibinfo {author}
  {\bibfnamefont {Y.}~\bibnamefont {Zhang}}, \ and\ \bibinfo {author}
  {\bibfnamefont {Z.}~\bibnamefont {Zheng}},\ }\href {\doibase
  10.1016/j.physa.2012.11.033} {\bibfield  {journal} {\bibinfo  {journal}
  {Physica A}\ }\textbf {\bibinfo {volume} {392}},\ \bibinfo {pages} {1232}
  (\bibinfo {year} {2013})}\BibitemShut {NoStop}%
\bibitem [{\citenamefont {Chen}\ \emph
  {et~al.}(2013{\natexlab{b}})\citenamefont {Chen}, \citenamefont {Nagler},
  \citenamefont {Cheng}, \citenamefont {Jin}, \citenamefont {Shen},
  \citenamefont {Zheng},\ and\ \citenamefont {D'Souza}}]{Chen13}%
  \BibitemOpen
  \bibfield  {author} {\bibinfo {author} {\bibfnamefont {W.}~\bibnamefont
  {Chen}}, \bibinfo {author} {\bibfnamefont {J.}~\bibnamefont {Nagler}},
  \bibinfo {author} {\bibfnamefont {X.}~\bibnamefont {Cheng}}, \bibinfo
  {author} {\bibfnamefont {X.}~\bibnamefont {Jin}}, \bibinfo {author}
  {\bibfnamefont {H.}~\bibnamefont {Shen}}, \bibinfo {author} {\bibfnamefont
  {Z.}~\bibnamefont {Zheng}}, \ and\ \bibinfo {author} {\bibfnamefont {R.~M.}\
  \bibnamefont {D'Souza}},\ }\href {\doibase 10.1103/PhysRevE.87.052130}
  {\bibfield  {journal} {\bibinfo  {journal} {Phys. Rev. E}\ }\textbf {\bibinfo
  {volume} {87}},\ \bibinfo {pages} {052130} (\bibinfo {year}
  {2013}{\natexlab{b}})}\BibitemShut {NoStop}%
\bibitem [{\citenamefont {G\'omez-Garde{\~n}es}\ \emph
  {et~al.}(2011)\citenamefont {G\'omez-Garde{\~n}es}, \citenamefont {G\'omez},
  \citenamefont {Arenas},\ and\ \citenamefont {Moreno}}]{Gomez-Gardenes11}%
  \BibitemOpen
  \bibfield  {author} {\bibinfo {author} {\bibfnamefont {J.}~\bibnamefont
  {G\'omez-Garde{\~n}es}}, \bibinfo {author} {\bibfnamefont {S.}~\bibnamefont
  {G\'omez}}, \bibinfo {author} {\bibfnamefont {A.}~\bibnamefont {Arenas}}, \
  and\ \bibinfo {author} {\bibfnamefont {Y.}~\bibnamefont {Moreno}},\ }\href
  {\doibase 10.1103/PhysRevLett.106.128701} {\bibfield  {journal} {\bibinfo
  {journal} {Phys. Rev. Lett.}\ }\textbf {\bibinfo {volume} {106}},\ \bibinfo
  {pages} {128701} (\bibinfo {year} {2011})}\BibitemShut {NoStop}%
\bibitem [{\citenamefont {Echenique}\ \emph {et~al.}(2005)\citenamefont
  {Echenique}, \citenamefont {{G\'omez-Garde\~nes}},\ and\ \citenamefont
  {Moreno}}]{Echenique05}%
  \BibitemOpen
  \bibfield  {author} {\bibinfo {author} {\bibfnamefont {P.}~\bibnamefont
  {Echenique}}, \bibinfo {author} {\bibfnamefont {J.}~\bibnamefont
  {{G\'omez-Garde\~nes}}}, \ and\ \bibinfo {author} {\bibfnamefont
  {Y.}~\bibnamefont {Moreno}},\ }\href@noop {} {\bibfield  {journal} {\bibinfo
  {journal} {Europhys. Lett.}\ }\textbf {\bibinfo {volume} {71}},\ \bibinfo
  {pages} {325} (\bibinfo {year} {2005})}\BibitemShut {NoStop}%
\bibitem [{\citenamefont {Knecht}\ \emph {et~al.}(2012)\citenamefont {Knecht},
  \citenamefont {Trump}, \citenamefont {{ben-Avraham}},\ and\ \citenamefont
  {Ziff}}]{Knecht11}%
  \BibitemOpen
  \bibfield  {author} {\bibinfo {author} {\bibfnamefont {C.~L.}\ \bibnamefont
  {Knecht}}, \bibinfo {author} {\bibfnamefont {W.}~\bibnamefont {Trump}},
  \bibinfo {author} {\bibfnamefont {D.}~\bibnamefont {{ben-Avraham}}}, \ and\
  \bibinfo {author} {\bibfnamefont {R.~M.}\ \bibnamefont {Ziff}},\ }\href
  {\doibase 10.1103/PhysRevLett.108.045703} {\bibfield  {journal} {\bibinfo
  {journal} {Phys. Rev. Lett.}\ }\textbf {\bibinfo {volume} {108}},\ \bibinfo
  {pages} {045703} (\bibinfo {year} {2012})}\BibitemShut {NoStop}%
\bibitem [{\citenamefont {Baek}\ and\ \citenamefont {Kim}(2012)}]{Baek11}%
  \BibitemOpen
  \bibfield  {author} {\bibinfo {author} {\bibfnamefont {S.~K.}\ \bibnamefont
  {Baek}}\ and\ \bibinfo {author} {\bibfnamefont {B.~J.}\ \bibnamefont {Kim}},\
  }\href {\doibase 10.1103/PhysRevE.85.032103} {\bibfield  {journal} {\bibinfo
  {journal} {Phys. Rev. E}\ }\textbf {\bibinfo {volume} {85}},\ \bibinfo
  {pages} {032103} (\bibinfo {year} {2012})}\BibitemShut {NoStop}%
\bibitem [{\citenamefont {Schrenk}\ \emph {et~al.}()\citenamefont {Schrenk},
  \citenamefont {Ara\'ujo}, \citenamefont {Ziff},\ and\ \citenamefont
  {Herrmann}}]{Schrenk14}%
  \BibitemOpen
  \bibfield  {author} {\bibinfo {author} {\bibfnamefont {K.~J.}\ \bibnamefont
  {Schrenk}}, \bibinfo {author} {\bibfnamefont {N.~A.~M.}\ \bibnamefont
  {Ara\'ujo}}, \bibinfo {author} {\bibfnamefont {R.~M.}\ \bibnamefont {Ziff}},
  \ and\ \bibinfo {author} {\bibfnamefont {H.~J.}\ \bibnamefont {Herrmann}},\
  }\href@noop {} {}\Eprint {http://arxiv.org/abs/arXiv:1403.2082v1}
  {arXiv:1403.2082v1} \BibitemShut {NoStop}%
\bibitem [{\citenamefont {Fehr}\ \emph {et~al.}(2009)\citenamefont {Fehr},
  \citenamefont {{Andrade Jr.}}, \citenamefont {{da Cunha}}, \citenamefont {{da
  Silva}}, \citenamefont {Herrmann}, \citenamefont {Kadau}, \citenamefont
  {Moukarzel},\ and\ \citenamefont {Oliveira}}]{Fehr09}%
  \BibitemOpen
  \bibfield  {author} {\bibinfo {author} {\bibfnamefont {E.}~\bibnamefont
  {Fehr}}, \bibinfo {author} {\bibfnamefont {J.~S.}\ \bibnamefont {{Andrade
  Jr.}}}, \bibinfo {author} {\bibfnamefont {S.~D.}\ \bibnamefont {{da Cunha}}},
  \bibinfo {author} {\bibfnamefont {L.~R.}\ \bibnamefont {{da Silva}}},
  \bibinfo {author} {\bibfnamefont {H.~J.}\ \bibnamefont {Herrmann}}, \bibinfo
  {author} {\bibfnamefont {D.}~\bibnamefont {Kadau}}, \bibinfo {author}
  {\bibfnamefont {C.~F.}\ \bibnamefont {Moukarzel}}, \ and\ \bibinfo {author}
  {\bibfnamefont {E.~A.}\ \bibnamefont {Oliveira}},\ }\href {\doibase
  10.1088/1742-5468/2009/09/P09007} {\bibfield  {journal} {\bibinfo  {journal}
  {J. Stat. Mech.}\ ,\ \bibinfo {pages} {P09007}} (\bibinfo {year}
  {2009})}\BibitemShut {NoStop}%
\bibitem [{\citenamefont {Fehr}\ \emph {et~al.}(2012)\citenamefont {Fehr},
  \citenamefont {Schrenk}, \citenamefont {Ara\'ujo}, \citenamefont {Kadau},
  \citenamefont {Grassberger}, \citenamefont {{Andrade Jr.}},\ and\
  \citenamefont {Herrmann}}]{Fehr11c}%
  \BibitemOpen
  \bibfield  {author} {\bibinfo {author} {\bibfnamefont {E.}~\bibnamefont
  {Fehr}}, \bibinfo {author} {\bibfnamefont {K.~J.}\ \bibnamefont {Schrenk}},
  \bibinfo {author} {\bibfnamefont {N.~A.~M.}\ \bibnamefont {Ara\'ujo}},
  \bibinfo {author} {\bibfnamefont {D.}~\bibnamefont {Kadau}}, \bibinfo
  {author} {\bibfnamefont {P.}~\bibnamefont {Grassberger}}, \bibinfo {author}
  {\bibfnamefont {J.~S.}\ \bibnamefont {{Andrade Jr.}}}, \ and\ \bibinfo
  {author} {\bibfnamefont {H.~J.}\ \bibnamefont {Herrmann}},\ }\href {\doibase
  10.1103/PhysRevE.86.011117} {\bibfield  {journal} {\bibinfo  {journal} {Phys.
  Rev. E}\ }\textbf {\bibinfo {volume} {86}},\ \bibinfo {pages} {011117}
  (\bibinfo {year} {2012})}\BibitemShut {NoStop}%
\bibitem [{\citenamefont {Daryaei}\ \emph {et~al.}(2012)\citenamefont
  {Daryaei}, \citenamefont {Ara\'ujo}, \citenamefont {Schrenk}, \citenamefont
  {Rouhani},\ and\ \citenamefont {Herrmann}}]{Daryaei12}%
  \BibitemOpen
  \bibfield  {author} {\bibinfo {author} {\bibfnamefont {E.}~\bibnamefont
  {Daryaei}}, \bibinfo {author} {\bibfnamefont {N.~A.~M.}\ \bibnamefont
  {Ara\'ujo}}, \bibinfo {author} {\bibfnamefont {K.~J.}\ \bibnamefont
  {Schrenk}}, \bibinfo {author} {\bibfnamefont {S.}~\bibnamefont {Rouhani}}, \
  and\ \bibinfo {author} {\bibfnamefont {H.~J.}\ \bibnamefont {Herrmann}},\
  }\href {\doibase 10.1103/PhysRevLett.109.218701} {\bibfield  {journal}
  {\bibinfo  {journal} {Phys. Rev. Lett.}\ }\textbf {\bibinfo {volume} {109}},\
  \bibinfo {pages} {218701} (\bibinfo {year} {2012})}\BibitemShut {NoStop}%
\bibitem [{\citenamefont {Fehr}\ \emph {et~al.}(2011)\citenamefont {Fehr},
  \citenamefont {Kadau}, \citenamefont {{Andrade Jr.}},\ and\ \citenamefont
  {Herrmann}}]{Fehr11}%
  \BibitemOpen
  \bibfield  {author} {\bibinfo {author} {\bibfnamefont {E.}~\bibnamefont
  {Fehr}}, \bibinfo {author} {\bibfnamefont {D.}~\bibnamefont {Kadau}},
  \bibinfo {author} {\bibfnamefont {J.~S.}\ \bibnamefont {{Andrade Jr.}}}, \
  and\ \bibinfo {author} {\bibfnamefont {H.~J.}\ \bibnamefont {Herrmann}},\
  }\href {\doibase 10.1103/PhysRevLett.106.048501} {\bibfield  {journal}
  {\bibinfo  {journal} {Phys. Rev. Lett.}\ }\textbf {\bibinfo {volume} {106}},\
  \bibinfo {pages} {048501} (\bibinfo {year} {2011})}\BibitemShut {NoStop}%
\bibitem [{\citenamefont {Schrenk}\ \emph
  {et~al.}(2012{\natexlab{c}})\citenamefont {Schrenk}, \citenamefont
  {Ara\'ujo},\ and\ \citenamefont {Herrmann}}]{Schrenk12b}%
  \BibitemOpen
  \bibfield  {author} {\bibinfo {author} {\bibfnamefont {K.~J.}\ \bibnamefont
  {Schrenk}}, \bibinfo {author} {\bibfnamefont {N.~A.~M.}\ \bibnamefont
  {Ara\'ujo}}, \ and\ \bibinfo {author} {\bibfnamefont {H.~J.}\ \bibnamefont
  {Herrmann}},\ }\href {\doibase 10.1038/srep00751} {\bibfield  {journal}
  {\bibinfo  {journal} {Sci. Rep.}\ }\textbf {\bibinfo {volume} {2}},\ \bibinfo
  {pages} {751} (\bibinfo {year} {2012}{\natexlab{c}})}\BibitemShut {NoStop}%
\bibitem [{\citenamefont {Schram}\ and\ \citenamefont
  {Barkema}(2012)}]{Schram12}%
  \BibitemOpen
  \bibfield  {author} {\bibinfo {author} {\bibfnamefont {R.~D.}\ \bibnamefont
  {Schram}}\ and\ \bibinfo {author} {\bibfnamefont {G.~T.}\ \bibnamefont
  {Barkema}},\ }\href {\doibase 10.1088/1742-5468/2012/03/P03009} {\bibfield
  {journal} {\bibinfo  {journal} {J. Stat. Mech.}\ ,\ \bibinfo {pages}
  {P03009}} (\bibinfo {year} {2012})}\BibitemShut {NoStop}%
\bibitem [{\citenamefont {Park}()}]{Park12}%
  \BibitemOpen
  \bibfield  {author} {\bibinfo {author} {\bibfnamefont {S.-C.}\ \bibnamefont
  {Park}},\ }\href@noop {} {}\Eprint {http://arxiv.org/abs/arXiv:1202.6024v2}
  {arXiv:1202.6024v2} \BibitemShut {NoStop}%
\bibitem [{\citenamefont {Marro}\ and\ \citenamefont
  {Dickman}(1998)}]{Marro98}%
  \BibitemOpen
  \bibfield  {author} {\bibinfo {author} {\bibfnamefont {J.}~\bibnamefont
  {Marro}}\ and\ \bibinfo {author} {\bibfnamefont {R.}~\bibnamefont
  {Dickman}},\ }\href@noop {} {\emph {\bibinfo {title} {Nonequilibrium {P}hase
  {T}ransitions in {L}attice {M}odels}}}\ (\bibinfo  {publisher} {Cambridge
  University Press},\ \bibinfo {address} {Cambridge},\ \bibinfo {year}
  {1998})\BibitemShut {NoStop}%
\bibitem [{\citenamefont {Park}(2013)}]{Park13}%
  \BibitemOpen
  \bibfield  {author} {\bibinfo {author} {\bibfnamefont {S.-C.}\ \bibnamefont
  {Park}},\ }\href {\doibase 10.3938/jkps.62.469} {\bibfield  {journal}
  {\bibinfo  {journal} {J. Korean Phys. Soc.}\ }\textbf {\bibinfo {volume}
  {62}},\ \bibinfo {pages} {469} (\bibinfo {year} {2013})}\BibitemShut
  {NoStop}%
\end{thebibliography}%

\end{document}